\documentclass[11pt, english]{article}
 
\usepackage[a4paper,
            bindingoffset=0.2in,
            left=0.8in,
            right=0.8in,
            top=0.8in,
            bottom=0.8in,
            footskip=.25in]{geometry}

\usepackage[colorlinks=true, citecolor=blue, allcolors=blue]{hyperref}
\usepackage[utf8]{inputenc}
\usepackage[T1]{fontenc}

\usepackage{mathtools}
\usepackage{graphicx}
\usepackage{float}
\usepackage{pdflscape}
\usepackage{color}
\usepackage[font=footnotesize,labelfont=bf]{caption}
\usepackage{amsthm}
\usepackage{amssymb}
\usepackage[colorinlistoftodos]{todonotes}
\usepackage{subfiles}

\usepackage{todonotes}
\usepackage{algorithm}
\usepackage{algpseudocode}
\usepackage[authordate, backend=biber]{biblatex-chicago}
\DeclareUnicodeCharacter{0301}{\'{e}}

\newenvironment{abstracttwo}{%
  \par\nobreak\noindent
  \textbf{\textit{Abstract}\hrulefill}\par\nobreak
  \small
  \noindent\ignorespaces
}{%
  \par\nobreak\normalsize
  \vskip-\ht\strutbox\noindent
  \textbf{\hrulefill}%
}

\title{Mapping Election Polarization and Competitiveness using Election Results}
\date{}

\usepackage{authblk}
\author[1,5]{Carlos Navarrete\thanks{Corresponding author: cnavarretel@udec.cl}}
\author[2,3]{Mariana Macedo}
\author[3,4]{Viktor Stojkoski}
\author[1]{Marcela Parada-Contzen}
\author[6]{Christopher Martínez}
\affil[1]{Departamento de Ingeniería Industrial, Facultad de Ingeniería, Universidad de Concepción}
\affil[2]{Northeastern University London}
\affil[3]{Center for Collective Learning, ANITI, Universit{\'e} de Toulouse}
\affil[4]{Department of Mathematics and Statistics, Faculty of Economics – Skopje,
Ss. Cyril and Methodius University in Skopje, Macedonia}
\affil[5]{Millennium Nucleus for the Study of Politics, Public Opinion and Media, Chile}
\affil[6]{Departamento de Administración Pública y Ciencia Política, Universidad de Concepción}

\renewcommand\Affilfont{\fontsize{9}{10.8}\itshape}

\DeclareCiteCommand{\autocite}[\mkbibparens]
  {\usebibmacro{prenote}}
  {\usebibmacro{citeindex}%
   \printtext[bibhyperref]{\usebibmacro{cite}}}
  {\multicitedelim}
  {\usebibmacro{postnote}}
  
\DeclareCiteCommand{\autocitep}[]
  {\usebibmacro{prenote}}
  {\usebibmacro{citeindex}%
   \printtext[bibhyperref]{\usebibmacro{cite}}}
  {\multicitedelim}
  {\usebibmacro{postnote}}

\begin{document}
\begin{refsection}
\maketitle

% \begin{center}
%     First draft: August 10, 2023 \\
%     Current draft: \today
% \end{center}

\begin{abstracttwo}

The simplified hypothesis that an election is polarized as an explanation of recent electoral outcomes worldwide is centered on perceptions of voting patterns rather than ideological data from the electorate. While the literature focuses on measuring polarization using ideological-like data from electoral studies--which are limited to economically advantageous countries and are representative mostly to national scales--we argue that, in fact, voting patterns can lead to mapping effective proxies of citizen divisions on election day. This paper perspectives two complementary concepts, Election Polarization (EP) and Election Competitiveness (EC), as a means to understand voting patterns on Election Day. We present an agnostic approach that relies solely on election data and validate it using synthetic and real-world election data across 13 countries in the Eurozone, North America, Latin America, and New Zealand. Overall, we find that we can label and distinguish expectations of polarized and competitive elections in these countries, and we report that EP positively correlates with a metric of political polarization in the U.S., unlocking opportunities for studies of polarization at the regional level and for lower/middle-income countries where electoral studies are available, but surveys are limited.

% Elections can unveil citizens' enthusiasm and discomfort concerning political candidates and parties. While a substantial literature studies elections from the perspective of winners and losers, we focused on an under-explored condition to unveil citizen divisions from voting patterns. This paper considers two complementary concepts: Election Polarization (EP) and Election Competitiveness (EC), which comprehensively measure the geography of citizen divisions on Election Day. We present an agnostic approach that relies on election data and considers candidates' competitiveness and geographical sorting. We validate the measures using synthetic and real-world election data across 13 countries in the Eurozone, North America, Latin America, and New Zealand, demonstrating that we can label and distinguish expectations of polarized and competitive elections. Our analysis reports that EP positively correlates with a metric of political polarization in the U.S., unlocking opportunities for studies of polarization at regional level and for lower/middle-income countries where electoral studies are available but surveys are limited.

\end{abstracttwo}

% \noindent Lorem ipsum dolor sit amet, consectetur adipiscing elit, sed do eiusmod tempor incididunt ut labore et dolore magna aliqua. 

% Introduction
% 2. Existent measures of Polarization and Competitiveness
% 3. A New Measure of Election Polarization and Competitiveness

% 2.1. Definition of Antagonism

% 2.1.1. Between-Antagonism
% 2.1.2. Within-Antagonism
% 2.1.3. Total Antagonism
% 2.2. Definition of Election Polarization
% 2.3. Definition of Competitiveness
% 2.4. Numerical Properties of Election Polarization and Competitiveness

% 3. Election Polarization and Competitiveness in Real Data
% 3.1. Data Source and Institutional Background
% 3.2. Descriptive Evidence
% 3.3. Empirical Specification of Analyses

% 4. Empirical Results

% 4.1. Understanding Election Polarization and Competitiveness
% 4.2. Political Polarization, Electoral Turnout, and Electoral Polarization
% 4.3. Robustness Check Across Aggregation Levels
% 4.4. Discussion

% 5. Conclusion

\section{Introduction}

Was the election polarized? This timely question, frequently overused by politicians and journalists after unforeseen electoral results, such as the two Chilean constitution referendums, the 2016-2020 U.S. presidential election, or the 2023 Argentinian election, has inspired recent work studying links between election results and polarization \autocite{lamont2017trump, layton2021demographic}. The salience of studying polarization relies upon its pivotal role in political processes. It can restrict the government's ability to respond to threats and overcome conflicts, potentially leading to democratic breakdowns \autocite{svolik2019polarization,perry2022american}. Although, it might help to mobilize the voters \autocite{ellger2023mobilizing}. Recently, a strand of literature argues that some neglected locations by globalization ``take revenge'' on big cities by voting for anti-establishment candidates \autocite{uscinski2021american,rodrik2018populism,rodriguez2018revenge} emerging a polarization that might not be related to preferences related to political proposals and competencies.

% Elections are the medium that citizens use to demonstrate their enthusiasm or discomfort concerning political candidates, parties, and issues \autocite{makarenko_2015}.

Throughout the last decade, political polarization has increased both in developed and developing countries. However, the literature lacks to capture modern polarization with empirical measures using open data \autocite{arbatli2021united}. While several definitions of polarization have been proposed (e.g., \autocitep{iyengar2012affect, fiorina2008political}), reliable data related to forms of political polarization, especially at the regional level, are expensive and challenging to collect. Overall, existing work depends on one-time electoral survey collected from a small amount of the population, which can only be done by a few established democracies as the United States \autocite{moral2023relationship,navarrete2023understanding,ryan2023exploring,hubscher2020does,arbatli2021united}, or by self-funded surveys \autocite{dancey2023personal,broockman2023does}. This limitation is significant because existing metrics, results, and implications are biased toward the developed world and cannot be generalized to other contexts. Hence, there is a need for cost-effective and simple metrics that can capture political polarization (or its proxies) also for less economically advantaged regions worldwide.

This paper proposes two complementary and agnostic metrics for mapping political polarization and electoral competition on Election Day using only observed electoral data. We call them Election Polarization (EP) and Election Competitiveness (EC). These measures complement each other and allow us to enhance our understanding of elections and polarization by disaggregating the footprint of voting patterns. In order to provide a simple metric that can be applied to virtually any election, we do not explicitly account for ideology in its conceptualization. Throughout this paper, we demonstrated that it is feasible to obtain a proxy for polarization relying only on voting patterns. Our definitions of the measures are inspired by the \autocitep{esteban1994measurement}'s identification-alienation framework and the social identity theory \autocite{tajfel1979integrative}. 
We argue that political candidates and parties in democratic countries encompass a pool of different ideological perspectives on society. Citizens who vote for the same candidate or political party feel more similar to each other than those with a different preference. For us, Election Polarization is a manifestation of a latent form of political polarization in the electorate. Political parties and candidates take positions and strategies to achieve their goals. This mobilization is neither spontaneous nor arbitrary: parties concentrate their efforts on places where they are more likely to succeed in their goals. These movements are reflected through voting patterns, which is a rich and openly available input that is accessible for virtually all elections in democratic countries.

Once the EP and EC measures are constructed, we compare their properties with traditional polarization and competitiveness measures in the literature. To validate these metrics towards a diverse overview of multiple electoral systems worldwide, we perform an extensive empirical analysis by collecting presidential and parliamentary election data for 13 countries. We collect data from the U.S., France, and Latin American countries such as Argentina, Brazil, Chile, Mexico, and Peru for presidential elections. We consider EU countries such as Belgium, Italy, Germany, and Spain for parliamentary elections. We also include Canada and New Zealand to consider systems that stem from British tradition. We do this by evaluating whether we can distinguish theoretical expectations of politically polarized or competitive elections and regions using the collected data. We then move to statistically assess the correlation across the EP and EC and the existent measures. Our analysis suggests that we can define a polarized election by only knowing whether candidates' preferences are geographically clusterized. To further validate the metrics, we perform a robustness check, demonstrating that our results can be generalized to any aggregation scale (e.g., city, county, state, or country), can be used for virtually any election type and electoral system (e.g., presidential or senate, majoritarian or proportional systems), are robust towards the number of candidates, the (effective) number of parties, and the level of abstentions. Altogether, these results provide an interesting setting for analyzing voters' behaviors across different nations. To our knowledge, this is one of the first efforts aiming to develop a comparable measure of polarization at regional and country levels. 

The rest of the paper is organized as follows. The next section discusses current measures of polarization and competitiveness. Section \ref{sec:metrics} presents the theoretical definition of the proposed measures on Election Polarization and Election Competitiveness. Section \ref{sec:results} presents theoretical and empirical properties of the measures. Section \ref{sec:future} perspectives on the usefulness of our metrics for future research, and, finally, Section \ref{sec:conclusion} concludes.

\section{Theoretical Discussion}

\textit{Is the population more polarized than in the past?} Likely, no other contemporary debate gets more attention from such diverse disciplines than this question. Nevertheless, defining polarization is not simple. Broadly, we define polarization as the division into two or more contrasting groups in a society. Related literature makes the distinction into several branches, such as {issues} \autocite{dimaggio1996have,navarrete2023understanding}, {geographical} \autocite{dalton2008quantity}, {socioeconomic} \autocite{muraoka2021does,hetherington2009putting,esteban1994measurement,wolfson1994inequalities}, {mass \& elite} \autocite{fiorina2008political,hetherington2001resurgent}, {public opinion} \autocite{kleiner2018public}, {partisan \autocite{druckman2013elite}}, or {affective} \autocite{dias2022nature,costa2021ideology,iyengar2019origins,iyengar2012affect}, to name a few. The salience of capturing its temporal and spatial dynamics relies on their effects on our society. 

Polarization is not a simple concept to model, as it is both a state and a process \autocite{dimaggio1996have}. Moreover, polarization affects groups differently \autocite{callander2022cause}. Regarding the causes of polarization, there are still gaps in the literature \autocite{mccarty2019polarization}. Empirical evidence indicates that an increase in inequality at the local level can be associated with an increase in the probability of supporting an extreme ideologically political party \autocite{winkler2019effect}. Partisan polarization is associated with greater importance for substantive policy considerations for citizens' vote choice in competitive elections \autocite{lachat2011electoral}. More recently, affective and elite polarization has been associated with lower attitudes towards COVID-19 \autocite{charron2022uncooperative}. Nonetheless, not all associations for polarization have a bad connotation. Previous work has established that political polarization is positively associated with control of corruption perception \autocite{testa2012polarization,brown2011political} and \autocitep{shi2019wisdom} found evidence that more ideologically polarized editorial teams produce articles of a higher quality than homogeneous teams.

To measure forms of polarization, related literature has relied on self-reported surveys \autocite{HomeANES47:online, waugh2009party}, social media data \autocite{barbera2015birds}, election data \autocite{arbatli2021united,johnston2016spatial, abramowitz2008polarization}, socio-demographic characteristics \autocite{scala2015red}, network analysis \autocite{vasconcelos2021segregation, hohmann2023quantifying}, individual preferences \autocite{faliszewski2023diversity}, to name a few. For instance, a line of argument in economics proposes measurements for polarization inspired by the shrinking of the middle-class' phenomenon, contrasting with existing metrics in inequality. A notable example is the seminal contribution of \autocitep{esteban1994measurement}. These authors aim to generalize polarization metrics upon a common framework, suggesting that society is polarized whether there are individuals who identify themselves with those having a similar feature (e.g., income) and are alienated from individuals in other groups (identification-alienation framework).

Despite the interest it is tracked in academic circles and has far-reaching real-life implications, there is an important missing puzzle piece in this literature: existing measures of polarization are not easily accessible and comparable to understanding its causes, conditions, and effects worldwide and at a regional scale. Data availability is biased toward the developed world, focused on recent years, and usually obtained via one-time data collected through expensive electoral studies and surveys (e.g., \autocitep{diermeier2023dynamics, dias2022nature,bubeck2022states,lee2021more}). While these studies offer a comprehensive overview of polarization, their economic cost makes it challenging to gather, at the same time, reliable individual ideological data and a representative view at regional levels. Furthermore, some simplified hypotheses that arise during electoral times, such as ``the population is more polarized than in the past'' and ``the elections are polarized,'' do not fit the traditional metrics of affective, partisan, or political polarization. Instead, these statements are derived from the voting patterns in the election itself and are grounded exclusively on the electoral results. Our approach addresses these limitations and contributes a more globally applicable perspective on polarization in electoral times.

\section{Our Measure}
\label{sec:metrics}

% Explain how the new measure builds upon the identification-alienation framework. Also, succinctly explain this framework to provide context.
Inspired by existing work using election results and building upon the identification-alienation framework of Esteban and Ray, we propose a simple and agnostic metric of polarization applicable worldwide only using election data. The main reason for pursuing this cost-effective approach lies in the fact that elections in democratic countries represent a rich and reliable data source on society's preferences. As transparency guarantees reliability in democracy, election data is often openly shared at disaggregated levels, and some widely-known democracy indexes such as the ``Democracy Index'' \autocite{kekic2007economist} use election outcomes to provide measures of democracy wealth.

Conceptually, Election Polarization (EP) manifests latent citizen divisions on election day. While we recognize that this conceptualization contrasts with the traditional view that polarization measures explicitly rely on individual ideological data, we argue that empirically, this data is challenging and expensive to collect, especially if we are interested in studying polarization dynamics at local levels and, most importantly, in economically disadvantaged countries without electoral studies tradition. At the core of our conceptualization lies the concept of geographical clustering (e.g., \autocitep{kim2003spatial}) and the principle of dispersion \autocitep{dimaggio1996have}. This dimension refers to the phenomenon in which the population with similar political preferences is concentrated in specific geographical areas. This involves the existence of electoral strongholds and targeted strategies of political candidates and parties. For us, grounded in the social identity theory and the identification-alienation framework, it is reasonable to argue that citizens who choose the same candidate (or party) should be, on average, more similar than those who prefer a different candidate, and an effective form of mapping this polarization is through the voting dispersion within a territory.

To construct our metric for election polarization, we account for the electoral competition as a complementary characteristic. While polarization and competitiveness are traditionally studied as different phenomena, we suggest that their interplay is essential to understanding the levels of polarization depicted in voting patterns. For instance, a two-candidate election in which each one holds 50\% of the preferences is not necessarily polarized. Both candidates could be just very competitive or undistinguished to voters. In this strand of literature, \autocitep{blais2009general} define a competitive election as the one in which its outcome is uncertain. Traditionally, the conventional measure of electoral competition is the margin of victory: the distance between the two leading candidates or parties. Existing literature proposes metrics of election competition based on district-level data taking into account the characteristics of the electoral system, voting share, and number of votes \autocite{grofman2009fully, blais2009general}, and they implicitly rely on assumptions concerning parties mobilizing power efforts to translate into votes \autocite{cox2020measuring}. For us, competitiveness must map the distance between all candidacies, no matter the electoral rule.

% However, the margin of victory is not a meaningful measure when referring to the diversity of electoral systems around the world (e.g., Proportional Representation (PR) systems or Single Member Elected (SMEs) with multiple parties). 

Our work implicitly assumes that voters identify themselves as similar to those who choose the same candidate and feel alienated from those with a different candidate or party preference. To capture polarization and competitiveness, we introduce a new concept applied to political candidates or parties: their antagonism.

\subsection{Definition of Antagonism}
\label{methods_antagonism}

% To the best of our knowledge, there is no agnostic method to map polarization by using electoral outcomes from any election type, such as councilors, primaries, or presidential, and for any number of candidates.

% To tie both concepts, we introduce the antagonism of candidates. 

We define the antagonism of a candidate $i$ as the divisiveness level that she or he generates in the electorate. This antagonism can be reflected by three characteristics: i) The competitiveness between candidates, ii) the voting dispersion of a candidate within a geography, and iii) her relative relevance measured by the number of votes. To better distinguish them, we decompose the antagonism by in-group (or \textbf{between-}) and out-group (or \textbf{within-}) components. We subsequently show that between-antagonism captures competitiveness and within-antagonism captures polarization.

First, we introduce some notation. Let $N$ be the number of (effective) candidates or parties, and let $M$ be the number of electoral units (e.g., district, polling stations). For a candidate $i$, let $share_{i,k}$ and $votes_{i,k}$ be her percent and number of votes in $k$-unit, respectively, and let $\overline{share_{i}}$ be the share of $i$ in the election. We now explore each component of antagonism.

\subsection{Within-Antagonism}

This dimension explores the individual performance of a candidate $i$ \textbf{within} a geography. In other terms, the existence of geographical clustering in preferences. Consider a city calling for candidacies to elect a future mayor. The city is divided into two districts: \textit{Alpha} and \textit{Beta}, composed of 300 voters each. In turn, the districts divide the electors into three precincts equitably. Let \{$A, B, C$\} be three candidates running for the position. By exploring the outcome per district (\textit{Alpha}: $A$=$30,29,31$; $B$=$20,19,21$; $C$=$50,52,48$ and \textit{Beta}: $A$=$5,0,85$; $B$=$5,40,15$; $C$=$90,60,0$--Each number represents a precinct within the district), we observe that candidates obtained the same votes ($C=150, A=90, B=60$). However, we observe that the voting dispersion for each candidate at the precinct level was almost null in \textit{Alpha} in comparison to \textit{Beta}, where candidate $C$ even obtained 0 votes in a precinct of \textit{Beta}. This idea leads to the definition of \textbf{Within-Antagonism} as a form to capture the voting dispersion of a candidate within a geography.

\begin{equation}
    \label{eq_dv_within}
    \text{Within-A}_{i} = 
        \frac{\sum\limits_{k=1}^{M}{votes_{i,k}|share_{i,k} - {\overline{share_{i}}}|}}
        {
        \left(N-1\right)\sum\limits_{k=1}^{M}{votes_{i,k}}
        }
\end{equation}

The \textbf{Within-Antagonism} ranges from 0 to $1/N$, where values closer to $1/N$ indicate that the distribution of the votes on candidate $i$ is less balanced. By construction, as this dimension of antagonism grounds on the dispersion of votes, we subsequently demonstrate that this can be equivalent to a proxy for political polarization.
% , we argue that it could capture rural-urban, north-south, or poor-rich groups' divisions, to name a few. In order to establish this link, we subsequently explore this statement. 

\subsection{Between-Antagonism}

This dimension relies on the competitiveness between candidates. In line with the literature on electoral competition, we consider that a population divided 50-50 in an election of two candidates must be more competitive than a 99-1 outcome since the voters are far away from a unanimous consensus over the winner. The same rule can be generalized for elections with more than two candidates (33.3\% for three, 25\% for four, etc.). Furthermore, we point out that a result of 50-49-1 between three candidates should be labeled as competitive since the third candidate is irrelevant to the outcome. 

We formally define the \textbf{Between-Antagonism} of candidate $i$ as the distance (measured by the percent of votes) with respect to all the other candidates in an election as follows:
% Thus, \textbf{between}-antagonism of candidate $i$ is defined as follows:

\begin{equation}
    \label{eq_dv_between}
    \text{Between-A}_{i} = 
        \frac{\sum\limits_{j=1}^{N}{\sum\limits_{k=1}^{M}{votes_{i,k}(1-|share_{i,k} - {share_{j,k}}|)}}}
        {
        N(N-1) \sum\limits_{k=1}^{M}{votes_{i,k}}
        }
\end{equation}

% \frac{M - 1}{M} \sum\limits_{k=1}^{N}{w_{i,k}}

% We introduced a variation of the alienation proposed by \citeauthor{esteban1994measurement} in order 
To capture the distance between groups, we subtract the distance between two candidates from the unit. That is, whether the distance between two candidates is minimum (two candidates with the same percent of votes), the antagonism is maximum (or 1). Conversely, whether the distance between two candidates is maximal (one candidate reaches the unanimity of votes), the antagonism is minimal (or 0). The values range from 0 and $1/N$, when values closer to $1/N$ mean that the candidate has a greater between-antagonism. Thus, \textbf{Between-Antagonism} of a candidate $i$ represents how much this candidate was preferred over the others.

% We test this assumption in the following sections.

\subsection{Definition of Election Polarization (EP)}

We motivated the idea that each candidate's antagonism represents a fraction of information over the level of polarization and competition of an election, and the sum of all antagonisms present in society must represent metrics of competitiveness and polarization.

First, we conceptualize Election Polarization (EP) as a proxy for political polarization, and it empirically measures the dispersion of votes across a geography. This metric is formally defined as the sum of within-antagonisms from all (effective) candidates on Election Day, i.e.,

\begin{align}
\text{EP} &= \sum_{i = 1}^{N} {\text{Within-A}_i}.
\end{align}

The EP ranges from 0 to 1, with values closer to the unit indicating greater polarization.

\subsection{Definition of Election Competitiveness (EC)}

Second, analogous to our definition of EP, we conceptualize Election Competitiveness (EC) as a metric of closeness from all candidates competing in the election. This metric is formally defined as the sum of between-antagonisms from all (effective) candidates on Election Day, i.e.,

\begin{align}
\text{EC} &= \sum_{i = 1}^{N} {\text{Between-A}_i}, \\
\end{align}

The EC ranges from 0 to 1, with values closer to the unit indicating greater competitiveness.

Both metrics can be applied to elections under any electoral system, regardless of the number of candidates or precincts/aggregation units. Nevertheless, these metrics can be only comparable to elections with the same number of candidates (See Appendix). As both metrics complement each other, their summation is at most 1. This is achieved in elections in which all candidates obtain the same voting percentage.
% To validate our approach, we will demonstrate that both dimensions are associated with their respective theoretical expectations, and their summation in maximal closeness or polarization must be 1.

% \section{Results}

% Here, we present our method to estimate election polarization and use this approach to explore its boundaries by using simulations and data from the presidential elections in the United States, France, and Chile.

\section{Empirical analysis} 
\label{sec:results}

\subsection{Properties of Election Polarization and Election Competitiveness}
\label{sec:properties}

\begin{figure}[h!]
    \centering
    \includegraphics[width=1\textwidth]{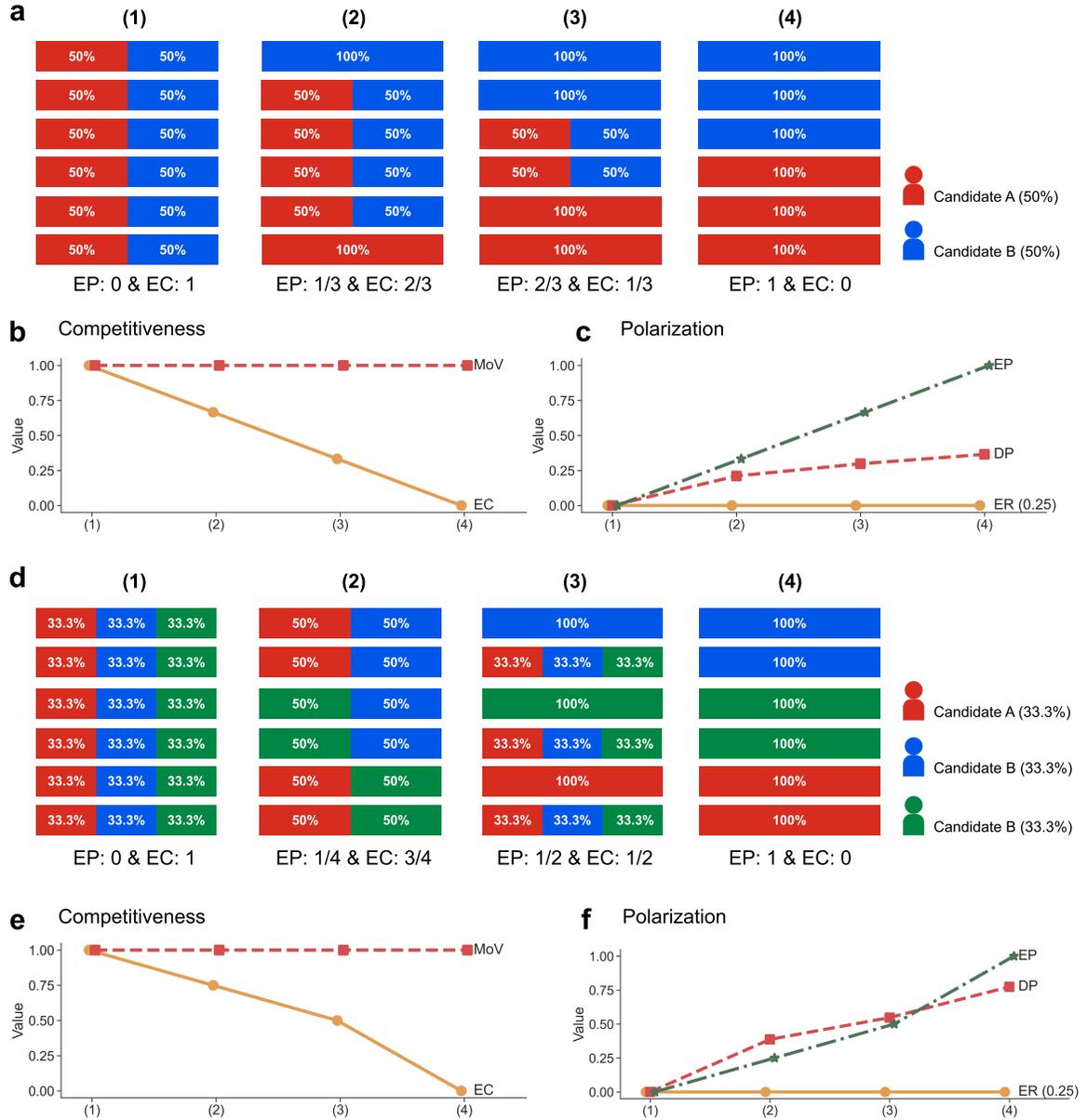}
    \caption{Edge-cases of eight fictional tied elections for two (a-c) and three (d-f) candidates. Each row represents a precinct, and each color represents a candidate. Comparison of traditional competitiveness (b, e) and polarization measurements (c, f) with our approaches for two (b-c) and three (e-f) candidates.}
    % \caption{\textbf{Numerical properties of Election Polarization}. (a) An illustrative example of our approach for two (Elections A-B) and three (Elections C-D) candidates. Each row represents a precinct, and each color represents a candidate. (b) Magic quadrant to summarize the criteria considered to map EP. X-axis represents the voting dispersion (associated with Within-EP), and the y-axis represents the closeness (associated with Between-EP). Violin plot distributions for synthetic election data for (c) two candidates and (d) three candidates. Each color represents a candidate. \textbf{Comparison of classical polarization measures for synthetic elections between two candidates}. (e-g) Correlation of polarization measures with the (e) EP, (f) Within-EP, and (g) Between-EP. Values in parenthesis for Esteban-Ray and Wang-Tsui represent the freedom degree used (see Appendix). We simulated 100 elections with 1000 units each, with an equal number of votes per unit (in our case, 100).}
    \label{fig:antagonism}
\end{figure}

We start exploring EP and EC numerically in order to understand their properties and draw comparisons with traditional measures of polarization and competitiveness. Figure~\ref{fig:antagonism} illustrates eight synthetic elections of (a-c) two and (d-f) three candidates with six precincts each--sorted from North to South. To exemplify edge cases of tied election results, we compare our EC and EP values with the ones calculated using Esteban-Ray (ER) and Dispersion (DP) for polarization and Margin of Victory (MV) for electoral competition to showcase their similarities and differences. While traditional measures for competitiveness and polarization treat all such elections as equals, our approach reveals hidden nuance by distinguishing whether the ties reflect close elections or rather geographical clustering of the electorate. 
% Notably, in our fictional tied elections, EP and EC always sum to 1, providing evidence of their complementarity. 

An additional advantage of our metrics can become apparent by looking at the correlation between EP and DP. While both metrics exhibit similar trends and successfully identify the most polarized election, as our metric ranges from 0 to 1, this feature enhances its interpretation.

We then expand the analysis by increasing the number of candidates. Expanding this observation through simulations (See Appendix), we find that in an election between $N$ candidates, maximal EP + EC is reached when the outcome of each candidate is closer to $1/N$ regardless of voting dispersion nor closeness, and minimal EP + EC is reached for unanimous elections. 
% ltogether, both features provide valuable information about election outcomes, contrasting with existing work that relies exclusively on the aggregate results.

\pagebreak 
\subsection{Election Polarization and Election Competitiveness Worldwide}
\label{sec:realdata}

% Next, we conducted an empirical study to explore whether we could quantify and identify theoretical expectations of politically ``polarized'' elections reported in the literature for Chile, France, and the United States. 

 % For this, we downloaded presidential election data from Chile (precinct level, 2013-2021) \autocite{Resultad22:online}, France (precinct level, 2002-2022) \autocite{Lesresul91:online}, and the United States (county level, 2000-2020) \autocite{baltz2022american}. These datasets were curated in order to be comparable over time (See Appendix for data cleaning procedure).

Next, we assess whether our metrics for EC and EP can identify and quantify theoretical expectations of politically polarized or competitive elections and regions. To this end, we perform an extensive empirical analysis using real-world election data by collecting presidential election data from Chile (2013-2021), Peru (2021), Brazil (2018-2022), Argentina (2023), Mexico (2018), France (2002-2022), and the United States (2000-2020) and parliamentary election data from Belgium (2014-2019), New Zealand (2023), Germany (2021), Italy (2022), Canada (2019), and Spain (2019) at the most disaggregated available level. 
% This data collection and curation effort also aims to provide construct validity to these metrics.
% {\color{red}A detailed set of infographics for each country and election can be found in the Appendix.}

\begin{figure}[H]
    \centering
    \includegraphics[width=1\columnwidth]{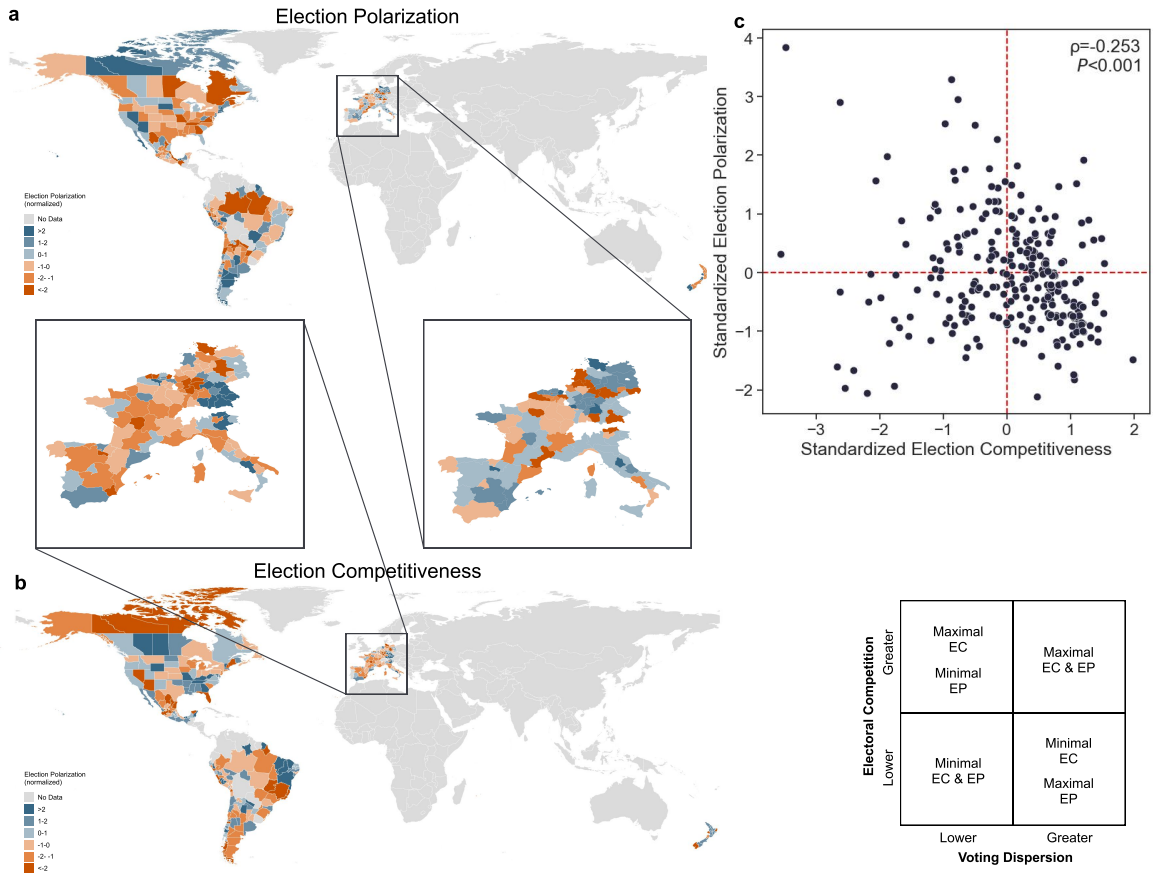}
    \caption{\textbf{The geography of Election Polarization and Competitiveness worldwide}. (a) Election Polarization and (b) Election Competitiveness. We calculated the regional-level values in Chile (2021), Argentina (2023), Brazil (2022), Peru (2021), Mexico (2018), the United States (2020), Canada (2021), France (2022), Spain (20), Belgium (2019), Germany (2021), Italy (2022) and New Zealand (2023). Values are standardized by subtracting their mean and dividing their standard deviation. (c) Relationship between standardized EP and EC.}

    \label{fig2}
\end{figure}

Figure \ref{fig2} reports the findings from polarized and competitive elections from 13 countries and 553 regions. To facilitate cross-regional comparisons, we normalize the values by subtracting the mean and dividing by the standard deviation per country. Furthermore, we report the empirical relationship of EC and EP in these regions. Figure \ref{fig2} a-b shows the election polarization and competitiveness, respectively. To highlight the potential of our metrics for successfully mapping theoretical expectations of polarized elections, we briefly mention some examples. We find that Catalonia is one of the most polarized regions in Spain, reflecting the tensions surrounding the Catalan independence movement \autocite{balcells2023secessionist}. Other notable cases reported in the figure are the polarization in the Southeast in the U.S., the clear cleavage between North and South in Brazil, and the peaks in regions with indigenous conflicts in the South of Chile. 
Next, Figure \ref{fig2} c reports the observed relationship of EC and EP. The correlation of both measures is rather low ($\rho=-0.253$, $P<0.001$). We find that, in general, most elections follow two main trends: the election can be classified as polarized, or rather it can be classified as competitive.

\begin{figure}[H]
    \centering
    \includegraphics[width=0.9\columnwidth]{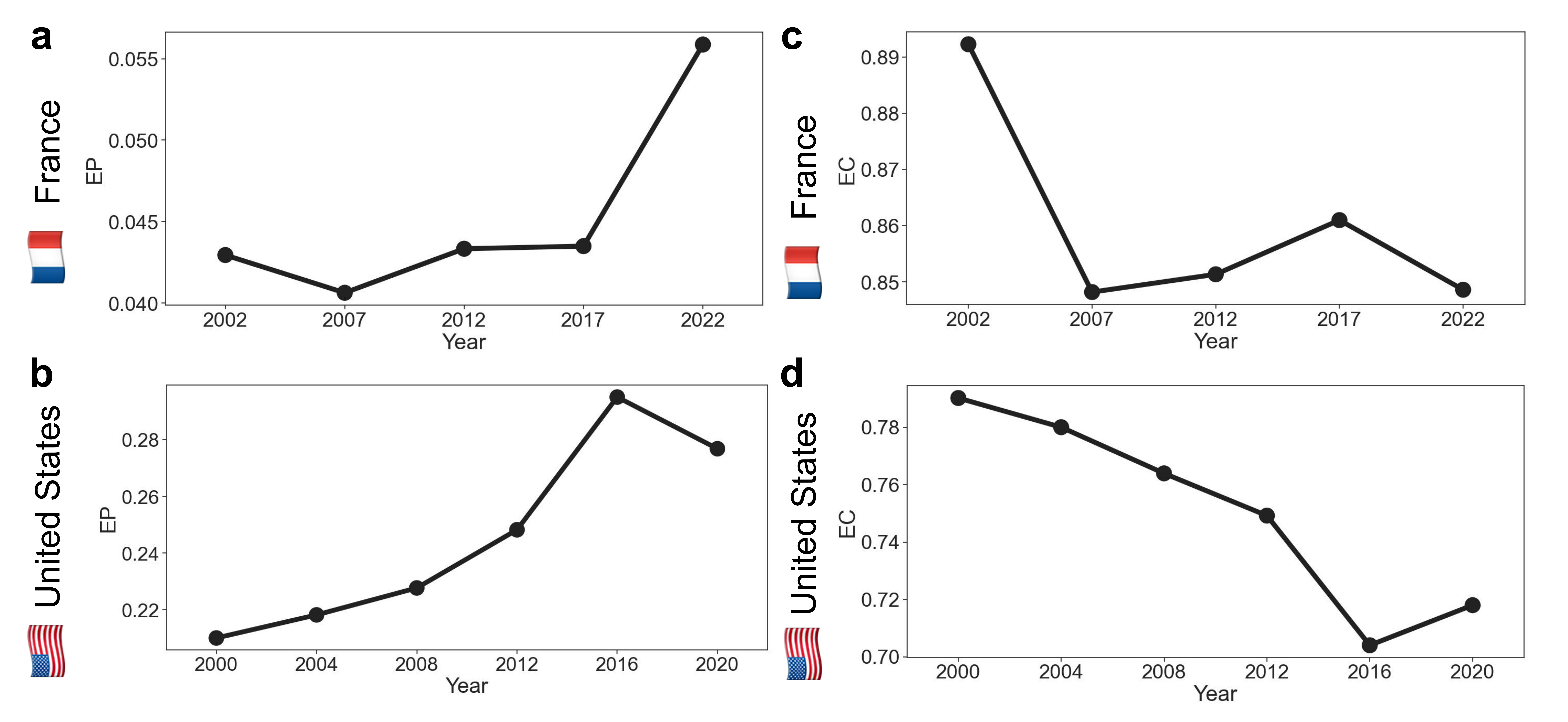}
    \caption{Temporal dynamics of Election Polarization in (a) France, and (b) the United States. Analogously, temporal dynamics of Election Competitiveness in (c) France and (d) the United States. In France, we used presidential election data from 2002 to 2022. In the United States, we used presidential election data from 2000 to 2020.}

    \label{fig3a}
\end{figure}

Then, we aggregate our analysis to explore the national-level dynamics of EP and EC in France and the United States over a 20-year period. We selected these countries since it is documented in previous work the existence of recent polarized elections. In the United States, Figure \ref{fig3a} a shows that this country experienced its most polarized election in 2016, followed by 2020. In France, we report the most polarized election in 2020, followed by 2017--with a 26\% increase among periods. Both results are aligned with the literature and the political climate in these countries. The Yellow Vest Movement in France \autocite{valentin2022polarization,chamorel2019macron} and the irruption of Donald Trump in the United States politics \autocite{abramowitz2019united} are well-known cases that have motivated recent polarization discussion in both countries. Following the previous results, we conducted an analogous analysis for competitiveness, with results shown in Figure \ref{fig3a} c-d. We found that the most competitive elections in the United States and France occurred in 2000 and 2002, respectively. Anecdotally, these peaks could be attributed to extremely competitive presidential elections. On the one hand, the 2000 election in the U.S. is well-known for the close result between Al Gore and George Bush (even less than 1000 votes difference in Florida) \autocite{Presiden94:online}, and the 2002 election in France was the first time that a far-right candidate made it into the run-off after a virtually close election among the three main majorities \autocite{mayer2013jean}. 

% Swing States

\begin{figure}[H]
    \centering
    \includegraphics[width=0.7\columnwidth]{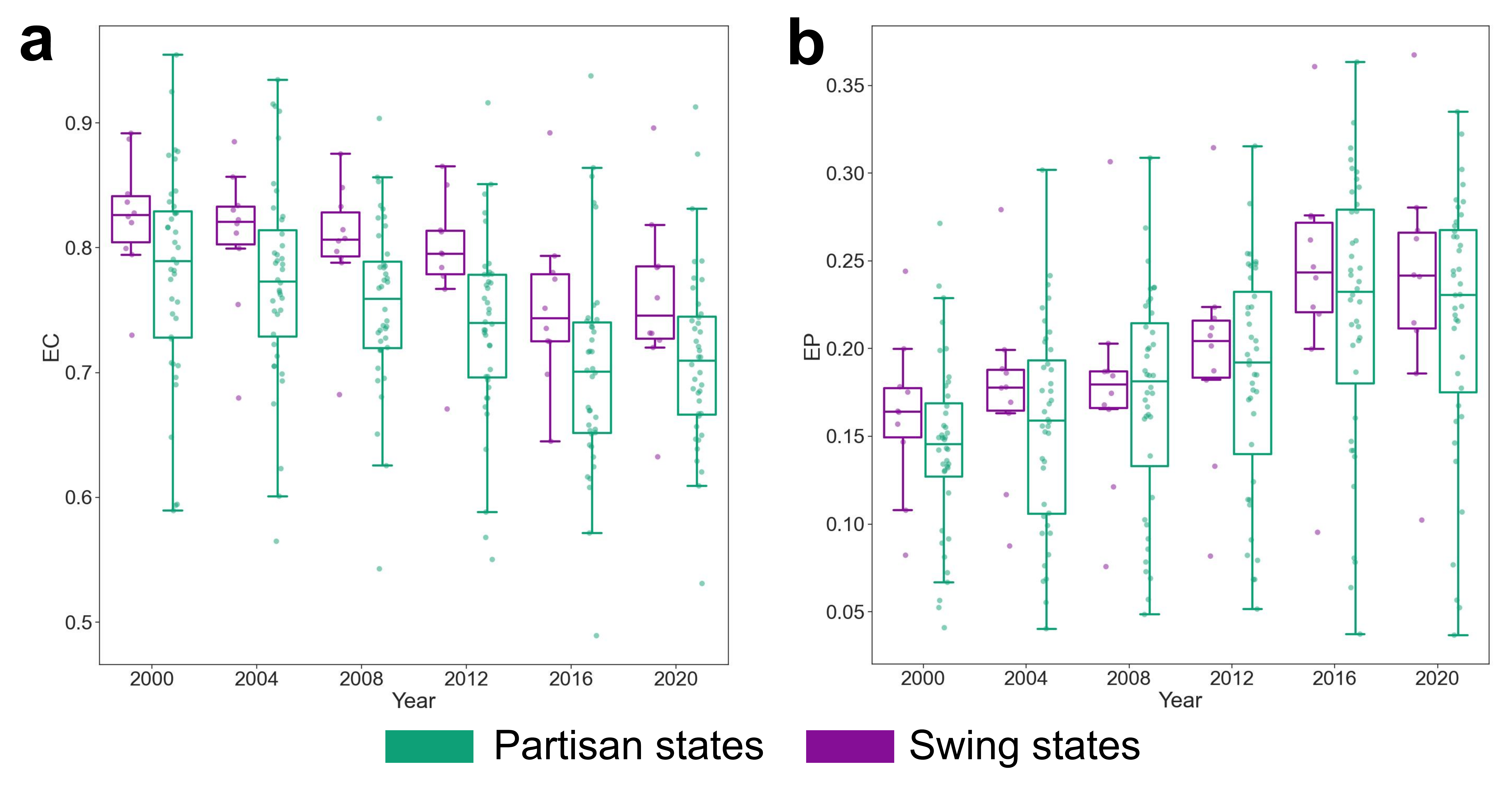}
    \caption{\textbf{Comparison of EP and EP in Swing and Partisan States in the United States}. (a) Election Polarization (b) Election Competitiveness. Each dot represents the value for a state.}

    \label{fig4a}
\end{figure}

Having established the feasibility and ability of our metrics to label theoretical expectations of polarization worldwide, we compare the EP and EC of Swing and Partisan states in the United States (Figure \ref{fig4a}). Swing states are considered crucial in the battle for the White House. In order to get elected, presidential candidates invest most of their campaign efforts in these battlegrounds \autocite{ahn2022three, Whyaresw57:online}, converting them into an excellent natural experiment to provide construct validity to our metrics (See Appendix). Our assumption considers that the elections must be competitive on average, but these battlegrounds should not be more polarized than other states. Figure \ref{fig4a} a finds that Swing states presented greater EC than Partisan states across time, and Figure \ref{fig4a} b finds these differences are explained mainly by EC rather than EP. In other words, we found more evidence to suggest that our methodological proposition for EC and EP captures different phenomena of elections.

\subsection{Election Polarization and Election Competitiveness and its Correlation with Mass Polarization and Electoral Turnout}

Sections \ref{sec:properties} and \ref{sec:realdata} offer insights demonstrating that our metrics for EP and EC discern levels of polarization from competitiveness in society. To further establish their usefulness, we now estimate regional-level political polarization and electoral turnout from traditional instruments to analyze their relationship with EP and EC. Our assumptions include that the political polarization around an election should be reflected on Election Day, and turnout must be positively associated with EC.

We focus on a state-level analysis in the United States since we ground on fine-grained data for a more extended study period.  We use data from the Cooperative Election Study (CES) \autocite{DVN/II2DB6_2022} and the U.S. Election Project \autocite{USElecti17:online}. Our interest is to investigate the combined characteristics that encompass political polarization--the ideological distance between Democrats and Republicans--and electoral turnout--proxied by the voting turnout on Election Day and to explore the correlation. Our analysis does not target causation.  Rather, we focus on understanding whether the dimensions of EP are related to political dimensions or whether socio-demographic features capture its effects. To ensure that socio-demographic characteristics and no other factors cause any increase in political polarization or turnout in the U.S., we control our analysis population, political, and economic characteristics. We also hold those unobserved factors in an Election Year by using year-fixed effects since an increase in polarization could result from multiple factors other than socio-demographic variables. For instance, the Cambridge Analytica scandal might play a role in the polarization identified in the 2016 Election. 

We set up panel regressions as follows:

\begin{align}
y_{g,t} &=  \beta_0\mathrm{EP}_{g,t} + \beta_1\mathrm{EC}_{g,t} \nonumber + \beta_2^ T\mathrm{X}_{g,t} + \beta_3\mathrm{S}_{g} + \mu_t + \epsilon_{g,t},
\label{eq:antagonism_model}
\end{align}

where $y_{g,t}$ is the dependent variable  (political polarization and electoral turnout) for the geography $g$ in year $t$), $\mathrm{EP}_{g,t}$ is the EP and $\mathrm{EC}_{g,t}$ is the EC. $\mathrm{X}_{g,t}$ is a vector of control variables that account for other factors addressed in the literature.  We control by inequality level (Gini coefficient), population density, education attainment, annual unemployment rate, and GDP per capita.
$\mathrm{S}_{g}$ is a dummy to control by swing states, $\mu_t$ is the period fixed effect, and $\epsilon_{g,t}$ is the error term. The variables have been normalized by subtracting the average and dividing by the standard deviation before running the regressions. Standard errors are clustered at the state level. The main results are presented in Figure \ref{fig4}. Additional models are presented in the Appendix.

\paragraph{Political Polarization}

In Figure \ref{fig4} (red line), we report correlations for political polarization with our EP and EC metrics, controlled by economic, socio-demographic, and political conditions. We find a positive relationship between EP and political polarization (0.31, $P=0.001$), though slightly less pronounced than educational attainment (-0.36, $P<0.001$). No significant association is identified for EC. Grounded on the results, we interpret that EP serves as a cost-effective proxy of political polarization, opening an avenue for more nuanced research on polarization at more granular levels, such as counties or districts. This is particularly salient for lower and middle-income countries lacking national surveys to characterize this phenomenon. 

\paragraph{Electoral Turnout}

In Figure \ref{fig4} (green line), we report correlations for electoral turnout with the EP and EC, conditional on economic, socio-demographic, and political conditions. We find that voting turnout positively correlates with EC (0.55, $P=0.003$), and it is the most important variable in terms of size effect in the model. A less significant association is observed for EP  (0.29, $P=0.036$), which, despite is not the main factor behind electoral turnout, its statistical significance provides valuable insights suggesting that more polarized elections lead to higher turnout. Since we additionally rely on voting turnout data for France and Chile, we perform a univariate correlation effect in both countries (France (NUTS-2):$\rho_{2022}=0.25$, Chile (region-level): $\rho_{2021}=0.19$), in line with theoretical expectations for a measure of electoral competition.

Going back to our initial question, our results suggest that election polarization can be a simple and suitable proxy for political polarization, and we set up our metric for election competitiveness as its complement.

\begin{figure}[h]
    \centering
    \includegraphics[width=0.85\textwidth]{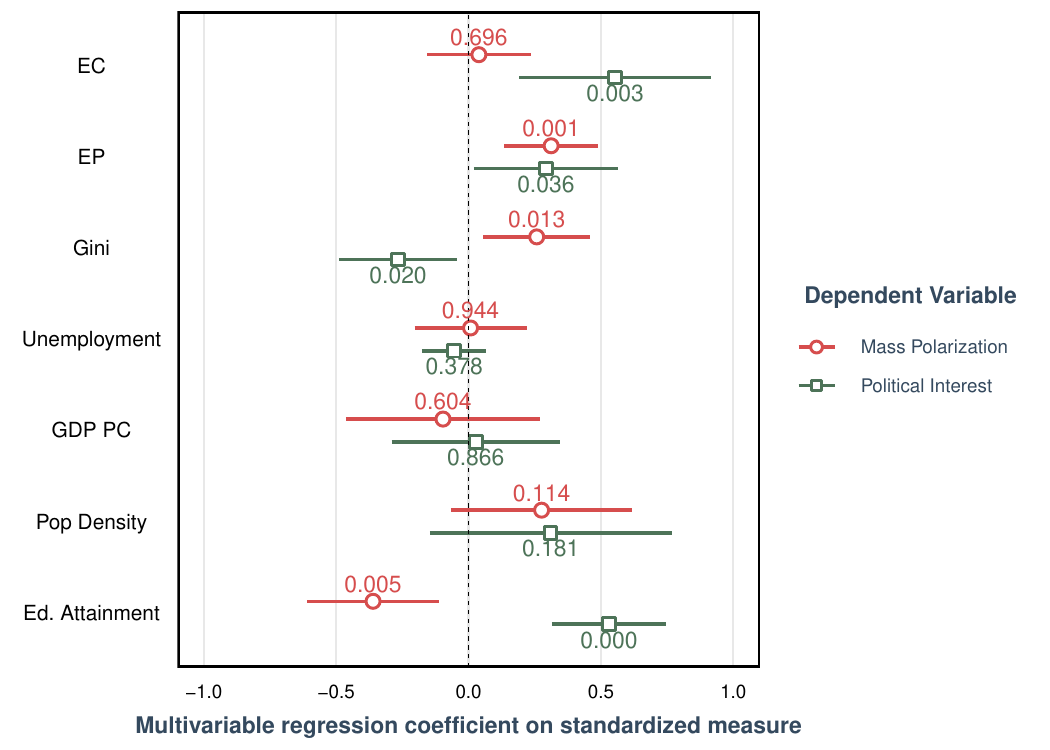}
    \caption{State-level Association of Mass Polarization and Electoral Turnout in the United States (2008-2020). Each model represents a multivariate regression, with year-fixed effects. Standard errors are in brackets and are robustly clustered by state. We standardized both independent and dependent variables to have a mean of zero and a standard deviation of one. Coefficients represent the p-values. $R^2_{\text{Mass Polarization}}=60\%$ and $R^2_{\text{Voting Turnout}}=62\%$. Additional Models can be found in the Appendix.}
    \label{fig4}
\end{figure}

\subsection{Robustness Check}

\begin{figure}[H]
    \centering
    \includegraphics[width=0.8\textwidth]{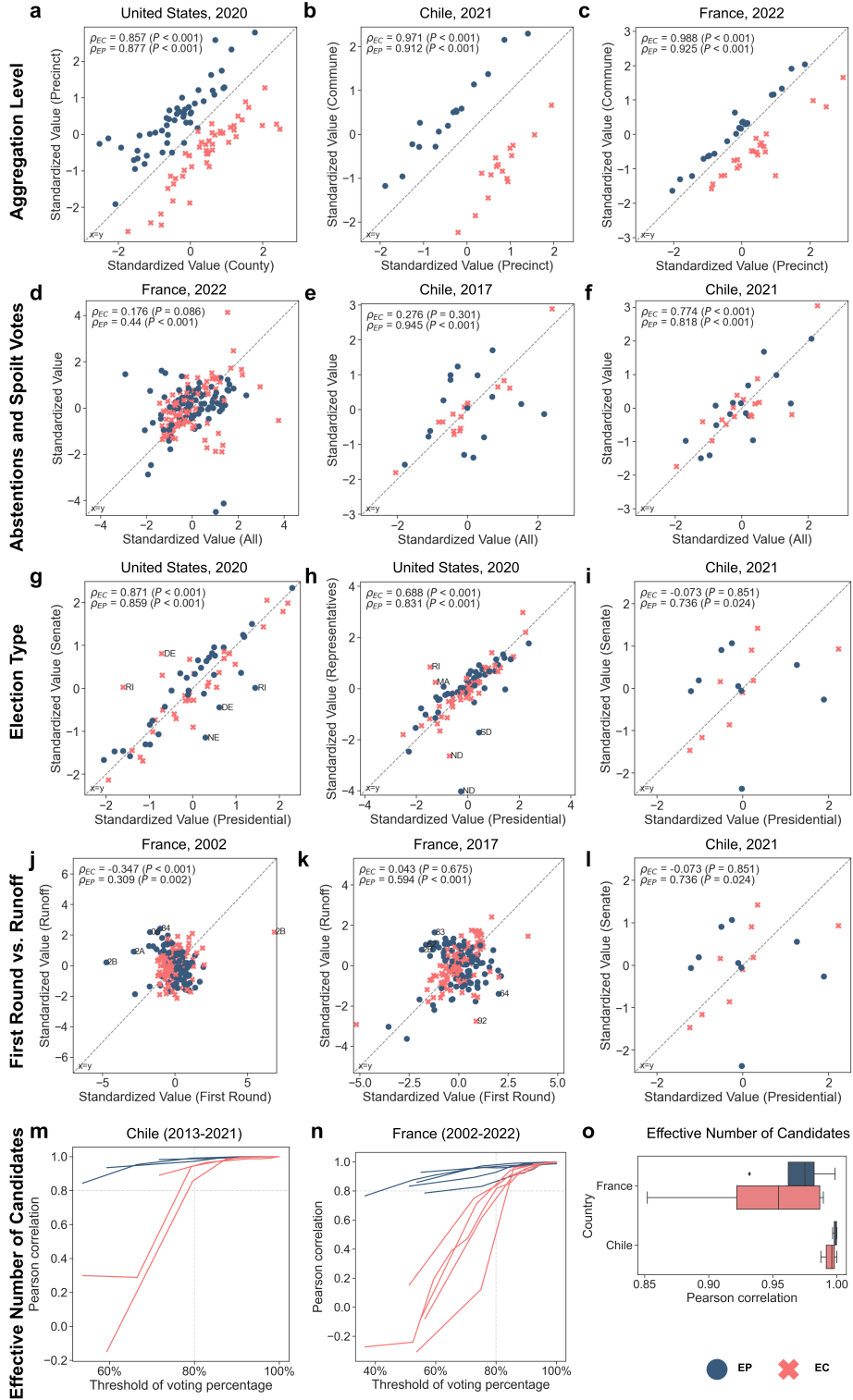}
    \caption{
    \textbf{Robustness Check using Real Data from Chile, France, and the United States.} Aggregation Level in the (a) United States, 2020; (b) Chile, 2021; and (c) France, 2022. Abstentions and Spoilt Votes in (d) France, 2022; (e) Chile, 2017; and (f) Chile, 2021. Election Type in (g) the United States, 2020, by comparing presidential and senate elections; (h) the United States, 2020, by comparing presidential and house elections; and (i) Chile, 2021, by comparing presidential and senate elections. First Round vs Runoff in (j) France, 2002; (k) France, 2017; (l) Chile, 2021. Effective Number of Candidates in (m) Chile, 2013-2021; (n) France (2002-2022); and (o) both countries. Values in $\rho$ represent the Pearson correlation. The variables have been normalized by subtracting the average and dividing by the standard deviation. The dashed lines from (a-l) represent $y=x$, and in (m-o) represent a correlation equal to 0.8.      }
    \label{fig:robustness}
\end{figure}

To conclude, we perform a robustness analysis to further validate and extend the reliability of our metrics by testing scenarios that could potentially bias its interpretation: i) the aggregation level, ii) abstentions and spoilt votes, iii) election type (i.e., two elections held on the same day), iv) the use of first round vs. runoff data, and v) the effective number of candidates (or parties). To this end, we compared variations of EP and EC calculated via these five approaches, and this analysis is reported in Figure \ref{fig:robustness}. We focus our analysis on Chile, France, and the United States, since we ground on fine-grained data to test these scenarios over a larger period.

\paragraph{Aggregation Levels}
First, we assess the use of different levels of granularity of data by comparing the regional-level EP and EC in France (department-level, 2022), the United States (state-level, 2020), and Chile (region-level, 2021). Discarding a potential bias toward the use of a specific aggregation level is relevant for elections in which data can not be obtained at polling station level. Specifically, we used precincts and county levels as our minimum aggregation unit. Figure \ref{fig:robustness} a-c report a strong positive correlation for both indices (France: EP$=0.925$, $P<0.01$, EC$=0.988$, $P<0.01$; United States: EP$=0.877$, $P<0.01$, EC$=0.857$, $P<0.01$; Chile: EP$=0.912$, $P<0.01$, EC$=0.971$, $P<0.01$), discarding the hypothesis that a specific aggregation level for our datasets altered the trends of our main results.

\paragraph{Abstentions and Spoilt Votes}

Second, we assess the level of abstentions, blanks, and spoilt votes in the election. To this end, we compared EP and EC computed using all the candidates with the one computed by including abstentions, spoilt, and blank votes treated as additional antagonists. We report a positive correlation for EP in Chile (Chile$_{2017}=0.945, P<0.001$; Chile$_{2021}=0.818, P<0.001$) and France (0.44, $P<0.001$). Nevertheless, this correlation is not statistically significant for EC (Chile$_{2017}=0.276, P=0.301$; Chile$_{2021}=0.774, P<0.001$, France$_{2022}=0.176, P=0.086$). Our analysis shows that overall, EP seems to be more robust than EC concerning levels of participation in elections, suggesting that abstentions might not bias the EP in most cases. We point out here that the statistical power is not maintained for all the cases, so this should be a concern while selecting the proper data to analyze. We infer that EC only reflects special election conditions, and additional analyses should be required to interpret it under scenarios with high abstention and spoilt votes. A potential non-causal explanation could be related to the fact that EC reflects a measure of competition exclusively in the election, but EP reflects a more profound condition of geographies, such as its polarization, as suggested in this study.

\paragraph{Election Type}
Third, we assess the EP and EC obtained from two elections held on the same day. In instances where multiple elections are celebrated simultaneously, we consider that citizens are subject to similar influences, such as discussions with friends, ad campaigns, TV, and social media. Consequently, those elections should reach a similar polarization level. To test this assumption, we use election data from general, senate, and representative elections in the United States in 2020 (precinct level) and senate and presidential elections in Chile in 2021. Figure \ref{fig:robustness} c-d shows a strong positive correlation between the EP computed for both elections (US$_{Representatives}$: EP$=0.831$, $P<0.01$, EC$=0.688$, $P<0.01$; US$_{Senate}$: EP$=0.859$, $P<0.01$, EC$=0.871$, $P<0.01$; Chile: EP$=0.736$, $P=0.024$, EC$=-0.073$, $P=0.851$). This finding enriches our metric for EP by suggesting its robustness and providing more evidence that EC can be interpreted as a measure applicable only to an election.

\paragraph{First round vs. Runoff}
Fourth, we compare the EP and EC obtained using first round and runoff presidential election data. Given that the time gap between both rounds is usually short, for example, two weeks in France and four weeks in Chile, it seems reasonable to think that changes in polarization would be marginal. Nevertheless, we know that candidates moderate their political positions in the second rounds \autocite{bordignon2016moderating} and citizens might be forced to vote by the ``lesser evil'' in some contests and not for their real preferences. Figure \ref{fig:robustness} reports the correlation of both indices for France (2002 and 2022) and Chile (2021) (More analyses in the Appendix). We systematically observe that the values from the first and second rounds are uncorrelated or exhibit a slight correlation. For instance, in the 2002 election in France, where it is documented that citizens voted against the far-right candidate, the correlation for EC was -0.347 ($P<0.001$), and the one for EP was 0.309 ($P=0.02$). To sum up, we conclude that the use of runoff data is not suitable for comparing the polarization or real competitiveness between rounds. Nevertheless, there is still a window of opportunity to study polarization in the same round from different years.

% Above a certain voting percentage, the EP should not vary significantly from the one obtained using all candidates. 
\paragraph{Effective Number of Candidates}
Finally, we evaluate whether our metrics calculated by excluding protest candidacies--or only focusing on the effective candidates \autocite{laakso1979effective}--can reach similar values to the ones using all candidates. This is relevant since the indices can be comparable only for elections with the same number of candidates over time. We perform two approaches: On the one hand, we compute EP and EC by relying on the top-2 candidates, then top-3, and so on, until completing all of them (See Appendix) by grouping the excluded candidates under a synthetic candidate called ``other''. On the other hand, we compare our metrics using all candidates and the one using the effective number of candidates. In both cases, we use presidential election data from France (2002-2022) and Chile (2013-2021), which feature more than two effective candidates per election. Figure \ref{fig:robustness} m-n reports the Pearson correlation of our indices computed from all candidates and the one calculated using the top-$N$ candidates, and Figure \ref{fig:robustness} o reports the Pearson correlation with respect to the second approach. In both analyses, we observe a similar pattern: We find a strong positive correlation (greater than 0.8) for any number of candidates that represent at least 80 \% of votes for both metrics, and surprisingly, EP can be mapped by only using the two main majorities in presidential systems, being robust to this finding for all elections analyzed. This analysis, however, could not be replicated for proportional systems, claiming for further analysis in this direction.

% \textcolor{red}{MM: But how much is from \% of votes, and how much is from the number of candidates?}

% Instead, comparing the polarization computed for 16 candidates in 2002 and 12 candidates in 2022 would bias the interpretation of the components.

These results computationally validate the usefulness of applying our metrics using election data under multiple assumptions (e.g., temporally, spatially) and relying on incomplete and noisy data, contributing to the reliability of this approach. From these analyses, we learn that EP reflects geographical phenomena, transcending the nature of election data and type. EC, nevertheless, can not be generalized to infer competitiveness in society beyond the scope of a particular election.

\section{New Measures’ Usefulness in Future Research}
\label{sec:future}

The finding that election polarization proxies political polarization has important implications for scholars. Our metrics can be applied to virtually any aggregation level, which highlights their potential to characterize polarization even at local levels, which was, thus far, virtually impossible to achieve. This means that EP can become an index of polarization of geographies, which is especially relevant for comparative studies worldwide. 

Furthermore, having demonstrated the feasibility of mapping our metrics using citizens' voting patterns on election day exclusively and its robustness under several scenarios and assumptions that could bias its interpretations, our perspective accounts for their usefulness as a starting point for studies in political science and economic geography. On the one hand, this study aligns with the active and timely debate on the determinants, conditions, and consequences of polarization worldwide, allowing us to leverage our knowledge even toward historical levels of polarization. On the other hand, scholars from economic geography interested in studying factors that predict the development of territories could use the EP to characterize how polarized regions could be associated with economic growth levels. 

This study did not analyze the antagonism of political candidates and parties. Analogous to the interpretation for EP and EC, we are able to obtain a metric of the political candidates' antagonism (or divisiveness) at virtually any aggregation level, allowing us to differentiate how divisive a candidate is throughout a territory. For instance, we observed that the antagonism of political candidates varies within a country; however, we did not explore the determinants and consequences of this unequal pattern. Our metric for antagonism can help study the chances of divisive candidates in runoff elections and also for the literature on median voters.

Altogether, these examples represent just a few avenues wherein our metrics can enrich our understanding of geographic dynamics and the extent of divisiveness and polarization within them. We expect to continue providing responses to fully understand the results of our methodological contribution.

\section{Conclusion}
\label{sec:conclusion}

This paper presents two complementary measures of Election Polarization and Competitiveness based on Election Day outcomes. These metrics make a valuable contribution to the existing literature by offering a straightforward and cost-effective proxy for delineating polarization. To construct the measures, we conceptualize the antagonism of political candidates and parties, which is represented by the closeness between candidates and voting dispersion across a territory. 

Importantly, this paper provides a diverse overview of worldwide electoral systems and political contexts. Using synthetic and empirical data from 13 countries, we demonstrated that our approach successfully labeled polarizing elections and growing trends presented in the literature, such as France, the United States, Chile, Spain, and Brazil. We demonstrated that EC and EP are straightforward metrics for understanding and comprehending polarization from a spatial perspective. Our approach has the potential to become an index of polarization in geographies, opening the possibility of extending the study of polarization beyond the U.S. borders and towards more granular scales. Nevertheless, we call for further focus on causation mechanisms between forms of ideological polarization and the voting patterns revealed on Election Day. A second stage could be linking the ideological position of candidates with the antagonism level and exploring the before and after election polls and their relationship with the EP. 

Our analysis reveals that Election Polarization is positively associated with a measure of political polarization, and  Election Competitiveness correlates to voting turnout. These results indicate that we can define a polarized election by simply knowing whether two conditions are satisfied: i) if there is total participation of the electorate (or close to that) in the election, and ii) if voters' preferences are highly clustered geographically. However, our findings do not claim causation, highlighting the need for additional behavioral studies to confirm that Election Polarization could be defined as a proxy for political polarization.

Demonstrating that election data can be a reliable proxy for mapping political polarization is especially relevant to the literature. Evidence from the 2018 Brazilian presidential election suggested that Jair Bolsonaro exploited demographic segregation to win the election \autocite{layton2021demographic}. Similar findings were observed in the 2016 United States presidential election, showing that unusually explicit appeals to racial and ethnic resentment and anti-establishment feelings attracted strong support from white working-class voters in favor of Donald Trump \autocite{abramowitz2019united}. Measuring and distinguishing polarization from election competition is essential to study their impacts on our society.

This study has some limitations that can be taken into consideration for further work. On the one hand, while our approach does not account for ideology in its conceptualization, we emphasized that if we want to provide a simple and representative form of mapping polarization at local scales, it is almost impossible to fulfill these conditions by including ideology in the equation. While some instruments provide ideological scores for political candidates or parties, we argue that this is not an attribute that can be obtained for all political parties/candidates worldwide. On the other hand, as the relationship between electorate sorting and political polarization has been established using electoral studies data from the United States, we call for further studies exploring the conditions and determinants of election polarization in different countries and its associations with sociodemographic conditions. One possibility is studying how the EP correlates to theoretical expectations of polarization, such as corruption control, inequality, or rurality, to name a few.

% , and the Within-EP is the dimension that mainly correlates with them. Interestingly, the state-level EP in the United States is positively associated with income inequality, which is consistent with the literature in the field \autocite{esteban2011linking,winkler2019effect}.

% Our analysis raises for additional  

% that get off polarization, voting patterns, and ideological polarization. 

% A second stage  %For example, the House of Representatives election outcomes in the United States do not work for computing state-level EP since citizens in two congressional districts from the same state cast different ballots. 
To conclude, we set up some considerations. First, the geographical hierarchy of an election must represent the most extensive aggregation level to calculate our metrics. The reason is that local candidates' quality could bias the party's support in a location \autocite{roy2015candidate}. Second, the EC is only valid in its corresponding election and can not be generalized to infer global levels of competition of parties/candidates. Third, our method is not valid for non-free-and-fair elections. To exemplify this, a referendum with 99\% approval in a dictatorship does not mean that society is not polarized at all. Finally, our metrics are unsuitable for comparing values between the first round vs runoff in the same electoral year since we find that, in some cases, both values are uncorrelated.

%Going forward, we expect that both EP and EC serve as a valuable insight to estimate the level of polarization in regions and contribute to policymakers and scholars looking for alternatives to mitigate it.

% \section*{Acknowledgments}

% We acknowledge the support and comments from participants at presentations of this work in Buenos Aires, Talca, Concepción, and Santiago. We also thank Professor Umberto Grandi for providing comments and suggestions on a preliminary version of this manuscript.

% \section*{Declaration of conflicting interests}
% The author(s) declared no potential conflicts of interest with respect to the research, authorship and/or publication of this article.

% \section*{Funding}
% The work of CN and MM was initially supported by Artificial and Natural Intelligence Toulouse Institute - Institut 3iA: ANR-19-PI3A-0004. The work of MP is supported by Intergenerational Mobility: From Modelling to Policy, Nucleo Millenio, ANID NCS2021\_072. The work of CN is supported by ANID – Millennium Science Initiative Program – NCS2021\_063. The funders had no role in study design, data collection and analysis, decision to publish, or preparation of the manuscript.

\section*{Data Availability Statement}
Upon the final acceptance of this paper, the data and source code used to perform this study will be publicly deposited on Harvard Dataverse.

%%%% Unnumbered equation

% \begin{acknowledgement}
% Insert the Acknowledgment text here.
% \end{acknowledgement}

% \paragraph{Funding Statement}

% This research was supported by grants from the <funder-name> <doi> (<award ID>); <funder-name> <doi> (<award ID>).

% \paragraph{Competing Interests}

% A statement about any financial, professional, contractual or personal relationships or situations that could be perceived to impact the presentation of the work --- or `None' if none exist.

%\endnote in some journals will behave like \footnote; and \printendnotes will not output anything. 
% \printendnotes

% Bibliography
\printbibliography
%[heading=bibintoc]
% 

% \appendix

% \section{Example Appendix Section}

% Lorem ipsum dolor sit amet, consectetur adipiscing elit, sed do eiusmod tempor incididunt ut labore et dolore magna aliqua. Lorem ipsum dolor sit amet, consectetur adipiscing elit, sed do eiusmod tempor incididunt ut labore et dolore magna aliqua. Lorem ipsum dolor sit amet, consectetur adipiscing elit, sed do eiusmod tempor incididunt ut labore et dolore magna aliqua. 
\end{refsection}
\clearpage
\begin{refsection}
\appendix

\maketitle

\tableofcontents

\newpage

\section{Supplementary Methods}

\subsection{Data Collection}
\label{data-collection-and-cleaning}
% analysis performed

This section presents our procedure for downloading, curating, and standardizing elections from each country included in this study.

% \subsubsection{}
% \paragraph{Data Source}
% \paragraph{Time Period}
% \paragraph{Aggregation Level}
% \paragraph{Special Considerations}

\subsubsection{Argentina}
\paragraph{Data Source} Datos Argentina, the official website of the Argentinian government to share open data \autocite{argentinadata}. Data is deposited by Dirección Nacional Electoral.

\paragraph{Period} 2019 and 2023 First Round' Presidential Elections.

\paragraph{Aggregation Level} Data obtained at polling station level. We use provinces as our territorial division.

\subsubsection{Belgium}
\paragraph{Data Source} Direction of Elections (Direction des Elections in French) \autocite{belgiumdata}.
\paragraph{Period} 2019 General Election.
\paragraph{Aggregation Level} Data obtained at the commune level. We use NUTS-2 as our territorial division.

\subsubsection{Brazil}
\paragraph{Data Source} Tribunal Superior Eleitoral \autocite{brazildata}
\paragraph{Period} 2018 and 2022 First Round' Presidential Elections.
\paragraph{Aggregation Level} We use states as our territorial division.

\subsubsection{Canada}
\paragraph{Data Source} Elections Canada \autocite{canadadata}.
\paragraph{Period} 2021 Federal Election (Lower Chamber).
\paragraph{Aggregation Level} Data obtained at polling station level. We use states as our territorial division.

\subsubsection{Chile}

\paragraph{Data Source}
The \textbf{Servicio Electoral} (SERVEL in its acronym in Spanish) \autocite{ServelSe99:online} \footnote{{https://www.servel.cl/resultados-en-excel-por-mesa-a-partir-del-ano-2012/}}. 

\paragraph{Period}
2013-2021 First Round and Runoff Presidential Elections.

\paragraph{Aggregation Level}
Data obtained at polling station level. We use administrative structure (regions and provinces) as our territorial division. The datasets were curated considering the most recent administrative structure established in 2017 \autocite{LeyChile88:online}. Thus, we perform the analysis using the administrative structure as of April 2023 of 16 regions (sorted from North to South): Arica y Parinacota, Tarapacá, Antofagasta, Atacama, Coquimbo, Valparaíso, Metropolitana, O'Higgins, Maule, Ñuble, Biobío, Araucanía, Los Ríos, Los Lagos, Aysen, and Magallanes.

\subsubsection{France} 
% https://www.data.gouv.fr/fr/pages/donnees-des-elections/
\paragraph{Data Source}
Le ministère de l'Interieur et des Outre-mer \autocite{francedata}. 

\paragraph{Period}
2002-2022 First Round and Runoff Presidential Elections.

\paragraph{Aggregation Level}
Data obtained at the polling station level. We use NUTS-2 and NUTS-3 as our territorial divisions.

\paragraph{Special Considerations}
The NUTS-3-level analysis encompassed 96 territories from Metropolitan France, after removing overseas territories. In Section ``Robustness Check'' of the main manuscript, we converted the 18 circumscriptions from the department of Paris (75) into communes. This fact is due to the fact that department 75 is just composed of one commune (75056), making it impossible to aggregate data for Paris at the commune level.

\subsubsection{Germany}
\paragraph{Data Source} Die Bundeswahlleiterin \autocite{germanydata}.
\paragraph{Period} 2021 General (Lower Chamber) Election.
\paragraph{Aggregation Level} Data obtained at polling station level. We use NUTS-2 v2021 as our territorial division.

% % % https://www.bundeswahlleiter.de/bundestagswahlen/2021/ergebnisse/opendata.html

\subsubsection{Italy}
\paragraph{Data Source} Ministero dell'Interno \autocite{italydata}.
\paragraph{Period} 2022 General Election (Chamber of Deputies), held on September 25th.
\paragraph{Aggregation Level} Data obtained at the commune level. We use NUTS-2 v2021 as our territorial division.

\subsubsection{Mexico}
\paragraph{Data Source} National Electoral Institute (Instituto Nacional Electoral in Spanish) \autocite{mexicodata}.
\paragraph{Period} 2018 Presidential Election.
\paragraph{Aggregation Level} Data obtained at polling station level. We use Federative Entity (state) as our territorial division.

\subsubsection{New Zealand}
\paragraph{Data Source} Electoral Commission \autocite{nzdata}.
\paragraph{Period} 2023 General Election (held on 14th October 2023)
\paragraph{Aggregation Level} Data obtained at voting place level. We use regions as our territorial division.
\paragraph{Special Considerations} We excluded the Māori electorate not associated with a particular region.

\subsubsection{Peru}
\paragraph{Data Source} National Office of Electoral Processes (Oficina Nacional de Procesos Electorales in Spanish) \autocite{perudata}.
\paragraph{Period} 2021 Presidential Election.
\paragraph{Aggregation Level} Data obtained at polling station level. We use regions as our territorial division.

\subsubsection{Spain}
\paragraph{Data Source} Ministerio del Interior \autocite{spaindata}. 
\paragraph{Period} 2019 General Election (held in November).
\paragraph{Aggregation Level} Data obtained at polling station level. We use NUTS-2 v2021 as our territorial division.
\paragraph{Special Considerations} Considering that the form of government in Spain is a parliamentary monarchy, we focus our analysis and download procedure to get data from congress elections. In this case, the political parties act as the equivalent of candidates.

% % \subsubsection{Romania}

% % The \textbf{Autoritatea Electorală Permanentă} is an autonomous institution in charge of organizing and conducting elections and referendums in the country. This institution makes available an interactive website with presidential elections data at the polling station level run in the period 1992-2014. In this case, we simultaneously automate our download procedure by retrieving data from each polling station.

% % \subsubsection{Spain}
% % % https://elecciones.lne.es/resultados-elecciones/generales/
% % % https://infoelectoral.interior.gob.es/opencms/es/elecciones-celebradas/area-de-descargas/
% % The \textbf{Minister of Internal Affairs} official website makes available results from elections and referendums since 1977 at polling station level. Considering that the form of government in Spain is a parliamentary monarchy, we focus our analysis and download procedure to get data from congress elections. In this case, the political parties act as the equivalent of candidates.

\subsubsection{United States}

\paragraph{Data Source}
The MIT Election Data + Science Lab (MEDSL) supports research on elections in the United States by sharing election outcome data for the Presidential, Senate, and House elections at different aggregation levels. 

\paragraph{Period}
2000-2020 Presidential Elections and 2016-2020 House and Senate Elections.

\paragraph{Aggregation Level}
Data obtained at the county level for presidential elections from 2000-2020 \autocite{DVN/VOQCHQ_2018} and at the precinct level for elections from 2016 to 2020 \autocite{DVN/LYWX3D_2018, DVN/NLTQAD_2018, DVN/ER9XTV_2022, DVN/VLGF2M_2022, DVN/JXPREB_2022}. We use states as our aggregation unit. 

\paragraph{Special Considerations}
The robustness analyses were conducted using the precinct-level data for the 2016 and 2020 House, Senate, and Presidential elections. Each state was labeled as Swing or Partisan. The District of Columbia was not considered in the robustness validation neither regression analysis. Due to the structure of the U.S. electoral system, we conduct all the analysis by just considering the two main majorities per election. Even though, in most cases, the main two majorities correspond to a candidate from the Republican and Democrat parties, there are some exceptions, as in the 2016 Senate election in California. We point out that we do not present all precinct-level findings since we do not have data available from a broader period than 2016-2020. Nevertheless, as demonstrated in the paper, our method is robust to use different aggregation levels.

\subsection{Data Curation}

The pipelines were coded in Python, mainly using Jupyter Notebook, Pandas v1.5.1 \autocite{mckinney2011pandas}, SciPy v1.9.3 \autocite{virtanen2020scipy}, and NumPy v1.23.4 \autocite{harris2020array}. The regression analyses were performed in R \autocite{leifeld2013texreg}. The visualizations were created using Seaborn v0.12.1, Matplotlib v3.6.2, and ggplot2. Each Notebook includes a step-by-step explanation of both data cleaning, standardization, and analysis. We will publicly deposit the source code and curated datasets into a repository on GitHub. 

\subsubsection{Data Files}

The data is deposited in the folder data\_curated. The filename structure follows the convention: \textbf{\{country\}\_\{year\}\_\{round\}.csv.gz}, where the parameter \textbf{country} represents the country (Chile, France, United States), \textbf{year} represents the election year, and \textbf{round} represents the electoral round (values accepted are first\_round, runoff, general, senate, and house). Each file is compressed in GZIP (Table \ref{tab:features_dataset_1_2} for details).

We point out that we create a unique identifier for each unit called \textit{polling\_id} (e.g., the precinct in the United States for the general election). This feature concatenates hierarchical aggregate levels according to the data origin. For example, the \textit{polling\_id}  joins state and county FIPS in the United States (2000), and the \textit{polling\_id} joins region, province, commune, and polling station name in Chile (2021).

We created a complementary file named \textbf{\{country\}\_\{year\}\_\{round\}\_location.csv.gz}, that contains metadata regarding the voting unit. Both files can be concatenated using \textit{polling\_id} feature.

\subsubsection{Pipelines}

The pipelines aim to allow researchers to replicate the findings presented in this paper. Roughly, we divide the pipelines in three main folders, as follows:

\begin{itemize}
    % \item Download\_\textbf{\{country\}}.ipynb: In case that the data is available through an API or external link, these files generate an output that can be accessed from the Pipeline file.
    \item Pipeline: Details the data cleaning procedure for each country (or election in some cases).
    \item Maps: Creates a set of maps or visualizations.
    \item Robustness: Presents a set of robustness tests and validations. Values accepted for type are Abstentions, Multicandidate, Validation, and Runoff.
\end{itemize}

\subsection{Sociodemographic Characteristics}

Sociodemographic characteristics included in our regression analyses, such as inequality level, annual unemployment rate, or median household income, were downloaded from official public institutions (e.g., Census Bureau) or repositories maintained by scholars or private institutions. Here, we briefly describe the data origin and curation procedure.

\subsubsection{United States}

% https://www.census.gov/data/tables/time-series/dec/historical-income-states.html

% https://fred.stlouisfed.org/release/tables?rid=330&eid=391443

\paragraph{Census Bureau}
\begin{itemize}
    \item Land area in Sq. Km. from State Area Measurements \autocite{StateAre33:online}.
    \item Population (Estimation) (2000-2020)
\end{itemize}

\paragraph{American Community Survey (ACS)} This survey is the premier source for detailed population and housing information in the U.S. \autocite{American89:online}. The Census Bureau releases every year a version of ACS, providing detailed household and population information.

\begin{itemize}
    \item Income inequality (Gini coefficient) (2010-2020) \autocite{Historic59:online}
\end{itemize}

% \paragraph{Frank-Sommeiller-Price Series} 

% Data collection of inequality measures maintained by Professor Mark W. Frank, Ph.D. Professor of Economics at SHSU \autocite{TheUSInc38:online}. He offers a comprehensive panel of annual state-level income inequality measures constructed from individual tax filing data available from the Internal Revenue Service. We downloaded the Income inequality (Gini coefficient) in the period 2000-2018.

% \paragraph{American's Health Rankings} This platform, supported by the United Health Foundation, provides a collection of datasets.

% \begin{itemize}
%     \item Annual unemployment rate (2000-2020)
% \end{itemize}

\paragraph{State Health Access Data Assistance Center (SHADAC)} 
This institution at the University of Minnesota provides a collection of statistics in order to help bridge the gap between policy and health issues \autocite{BulkData55:online}. From SHADAC, we downloaded the following data sets:

\begin{itemize}
    % \item Income inequality (2006-2020) (Gini coefficient)
    \item Annual unemployment rate (2000-2020)
\end{itemize}

% https://statehealthcompare.shadac.org/Bulk#1,2,3,4,5,6,7,8,9,10,11,12,13,14,15,16,17,18,19,20,21,22,23,24,25,26,27,28,29,30,31,32,33,34,35,36,37,38,39,40,41,42,43,44,45,46,47,48,49,50,51,52/148,82

\paragraph{Bureau of Economic Analysis (BEA)}

The Bureau of Economic Analysis (BEA) is the federal agency responsible for producing statistics and data about the U.S. economy \autocite{Interact14:online}. We downloaded the following state-level data sets from BEA:

\begin{itemize}
    \item Personal Income per capita (in Dollars) (2000-2020).
    \item Real Gross Domestic Product (Millions of chained 2012 dollars) (2000-2020).
    \item Personal Expenditure per capita (2000-2020).
    % \item Population (2000-2020).
\end{itemize}

\subsubsection{Special considerations in the U.S.}

\paragraph{Income inequality} 
% As seen, there is no data source that includes the state-level Gini coefficient for all our period of study. We then followed two approaches: Firstly, we input the Frank-series Gini coefficient for 2018 as the one in 2020. 
We tested our models in a shorter period using the Gini from the Census and imputing 2010 as the value for 2008.

\paragraph{Income inequality}

We relied on the region-level Gini coefficient (using 2017 data) computed by Brevis (2020) \autocite{mieres2020dinamica}.

% We proxied those values in order to be compared with the last presidential election (in our case, 2021).

\subsection{Political Interest}
We proxy political interest as the participation rate on Election Day. From France (2022) and Chile (2017, 2021), election data already includes the number of abstentions per polling station. In the United States, we downloaded voting statistics from the U.S. Elections Project (2000-2020) \autocite{USElecti17:online}, and used the Voting Elegible Population (VEP) turnout rate. The univariate correlations can be found in Figure \ref{fig:corrturnout}.

\subsection{Synthetic data}
% random.gauss(mu=mu, sigma=sigma) for x in range(100)

Throughout the manuscript, we explored the numerical properties of EP using synthetic data. Algorithms \ref{alg:two_candidates} and \ref{alg:three_candidates} summarize our approach for elections of two and three candidates, respectively. Here we sampled $\mu \in \{0.5, 0.66, 0.75, 0.8333, 1\}$, $\sigma \in \{0.0025, 0.05, 0.10, 0.25\}$, and we set the number of units $M$ in 100. It should be noted that we consider as constant the number of votes per unit, in our case, 100.

\begin{algorithm}[hbt!]
\caption{Data sampling algorithm for a given election between two candidates. }
\label{alg:two_candidates}
\begin{algorithmic}
\Require $\mu \in [0,1], \sigma \in [0,0.25], m > 0, N \gets 2$
\Ensure $\sum\limits_{k=0}^1 r_{i,k} = 1 \And \sum\limits_{i=0}^{m-1}\sum\limits_{j=0}^1 v_{i,k} = 100m, \forall i \in [0, m-1]$

\State $a \gets \mu$ \Comment{Mean}
\State $s \gets \sigma$ \Comment{Standard deviation}
\State $M \gets m$ \Comment{Number of voting units}

\For{$i \gets 0$ to $M - 1$}
    \State $r_{0,i} \gets \mathcal{N}(a,\,s)$
    \If{$r_{0,i} > 1$}
        \State $r_{0,i} \gets 1$
    \ElsIf{$r_{0,i} < 0$}
        \State $r_{0,i} \gets 0$
    \EndIf
    
    \State $r_{0,i} \gets r_{0,i} - r_{0,i}\mod{0.01}$
    \State $r_{1,i} \gets 1 - r_{0,i}$
\EndFor

\For{$i \gets 0$ to $N - 1$}
    \For{$j \gets 0$ to $M - 1$}
    \State $v_{i,k} \gets 100r_{i,k}$
    \EndFor
\EndFor

\end{algorithmic}
\end{algorithm}

\begin{algorithm}[hbt!]
\caption{Data sampling algorithm for a given election between three candidates. }
\label{alg:three_candidates}
\begin{algorithmic}
\Require $\mu_1,\mu_2 \in [0,1], \sigma_1,\sigma_2 \in [0,0.25], m > 0, N \gets 3$
\Ensure $\sum\limits_{j=0}^2 r_{i,k} = 1 \And \sum\limits_{i=0}^{m-1}\sum\limits_{j=0}^2 v_{i,k} = 100m, \forall i \in [0, m-1]$

\State $a_1 \gets \mu_1$ \Comment{Mean (1)}
\State $a_2 \gets \mu_2$ \Comment{Mean (2)}
\State $s_1 \gets \sigma_1$ \Comment{Standard deviation (1)}
\State $s_2 \gets \sigma_2$ \Comment{Standard deviation (2)}
\State $M \gets m$ \Comment{Number of voting units}

\For{$i \gets 0$ to $M - 1$}
    \State $r_{0,i} \gets \mathcal{N}(a_1,\,s_1)$
    \State $r_{1,i} \gets \mathcal{N}(a_2,\,s_2)$
    \For{$j \gets 0$ to $1$}
         \If{$r_{i,k} > 1$}
            \State $r_{i,k} \gets 1$
        \ElsIf{$r_{0,i} < 0$}
            \State $r_{i,k} \gets 0$
        \EndIf
    \EndFor

    \If{$(r_{0,i} + r_{1,i}) > 1$}
        \State $r_{1,i} \gets 1 - r_{0,i}$
    \EndIf

    \State $r_{2,i} \gets 1 - (r_{0,i} + r_{1,i})$
\EndFor

\For{$i \gets 0$ to $M - 1$}
    \For{$j \gets 0$ to $N - 1$}
    \State $v_{i,k} \gets 100r_{i,k}$
    \EndFor
\EndFor

\end{algorithmic}
\end{algorithm}

\begin{algorithm}
\caption{An algorithm with caption}\label{alg:cap}
\begin{algorithmic}
\Require $n \geq 0$
\Ensure $y = x^n$
\State $y \gets 1$
\State $X \gets x$
\State $N \gets n$
\While{$N \neq 0$}
\If{$N$ is even}
    \State $X \gets X \times X$
    \State $N \gets \frac{N}{2}$  \Comment{This is a comment}
\ElsIf{$N$ is odd}
    \State $y \gets y \times X$
    \State $N \gets N - 1$
\EndIf
\EndWhile
\end{algorithmic}
\end{algorithm}

\subsection{Data Analysis}
\label{data_analysis}

We detail our data analysis procedure. Roughly, the file {\color{blue}\textit{pipeline\_files.py}} contains a function that converts the preprocessed election data into a measure of polarization and {\color{blue}\textit{pipeline.py}} is a simplified version to calculate EP and EC. This procedure can be summarized as follows:

\begin{itemize}
    \item We select an aggregation level. 
    \item In case of existence in the data, we exclude blank votes, spoilt votes, and abstentions.
    \item To compare elections with a different number of candidates over time, we maintain the top-2 majorities in the United States, four majorities in Chile, and eight majorities in France. The rest of the candidates were excluded from the data.
    \item We re-calculate candidates' voting percentages in each aggregation unit.
    \item We calculate Within- and Between- antagonism of $i$.
\end{itemize}

\subsection{Comparison of Polarization Measures}

We adapted traditional polarization measurements, such as Esteban-Ray and Dispersion, in order to compute alternative forms of antagonism level for candidate $i$ by considering the $M$ voting units. The algorithmic implementations of each method can be found in {\color{blue}\textit{Comparison\_Polarization\_Measures.ipynb}}.

\subsubsection{Esteban and Ray, 1994}

\begin{equation}
    ER_{i} = K \sum\limits_{k=1}^{M} \sum\limits_{j=1}^{N} v_{i,k}^{1+\alpha}v_{j,k}|r_{i,k}-r_{j,k}|
\end{equation}

Where $v_{i,k}$ is the number of votes of $i$ in $k$, $r_{i}$ the percent of votes of $i$ in $k$, and $K$ and $\alpha$ represents freedom degrees introduced by the authors. We set $\alpha = 0.25$ and $\alpha = 1$. In all cases, we set $K=\frac{1}{\left(\sum\limits_{k=1}^{M}v_{i,k}\right)^{(2 + \alpha)}}$

% \subsubsection{Wolfson, 1994}

% \begin{equation}
%     W_{i}=\frac{\mu_i}{m_i}2(2T-\text{Gini})
% \end{equation}

% Where $T=\frac{1}{2}-L(\frac{1}{2})$ and $L(\frac{1}{2})$ are the income share of the bottom half of the population, $m_i$ is the median percent of votes, $\mu_i$ is the mean percent of votes.

% \subsubsection{Wang and Tsui, 1998}

% \begin{equation}
%     WT_{i}=\frac{K}{P}\sum\limits_{k=1}^{M}v_{i,k}\left|\frac{r_{i,k}-m_i}{m_i}\right|^\gamma
% \end{equation}

% Where $P$ is the number of voters $P=\sum\limits_{k=1}^{M}v_{i,k}$, $m$ is the median percent of votes of $i$, $\gamma$ is a freedom degree that $\gamma \in \{0,1\}$, and $K$ also represents a freedom degree. We set $K=1$ and test with $\gamma=0.5$ and $\gamma=0.75$.

\subsubsection{Dispersion}
\begin{equation}
    \text{Dispersion}_{i} = \sqrt{\frac{\sum\limits_{k=1}^{M}\left(r_{i,k}-\mu_i\right)^2}{M-1}}
\end{equation}

Where $M$ is the number of units, $r_{i,k}$ the percent of votes of $i$ in $k$, and $\mu_i$ is the voting percentage of $i$ in the election.

\subsubsection{Polarization: Esteban-Ray, Wang-Tsui, Dispersion}

Let $A'$ be the \textbf{antagonism} of a candidate $i$ computed with a classical approach. As introduced in the manuscript, the total election polarization of a society is represented as follows:

\begin{equation}
    EP' = \sum\limits_{i=1}^{N}A'_i
\end{equation}

% \subsubsection{Polarization: Reynal-Querol (2002)}

% We also find another approach to measure polarization introduced by Reynal-Querol, which considers that a measure of polarization should be maximal if there are two groups of the same size in society.

% \begin{equation}
%     RQ = 1 - \sum\limits_{i=1}^{N}\left(\frac{1/2 - \mu_i}{1/2}\right)^2\mu_i
% \end{equation}

% Where $N$ is the number of candidates and $\mu_i$ is the voting percentage of $i$ in the election. It should be noted that RQ does not consider the voting dispersion throughout a territory for a candidate, and is motivated to capture religious polarization.

% % \subsubsection{Weighted Dispersion}
% % \begin{equation}
% %     \text{Weighted.Dispersion}_{i} = \sqrt{\frac{\sum\limits_{k=1}^{M}v_{i,k}\left(r_{i,j}-\mu_i\right)^2}{ \frac{M - 1}{M}\sum\limits_{k=1}^{M}v_{i,k}}}
% % \end{equation}

\begin{figure}[H]
    \centering
    \includegraphics[width=\columnwidth]{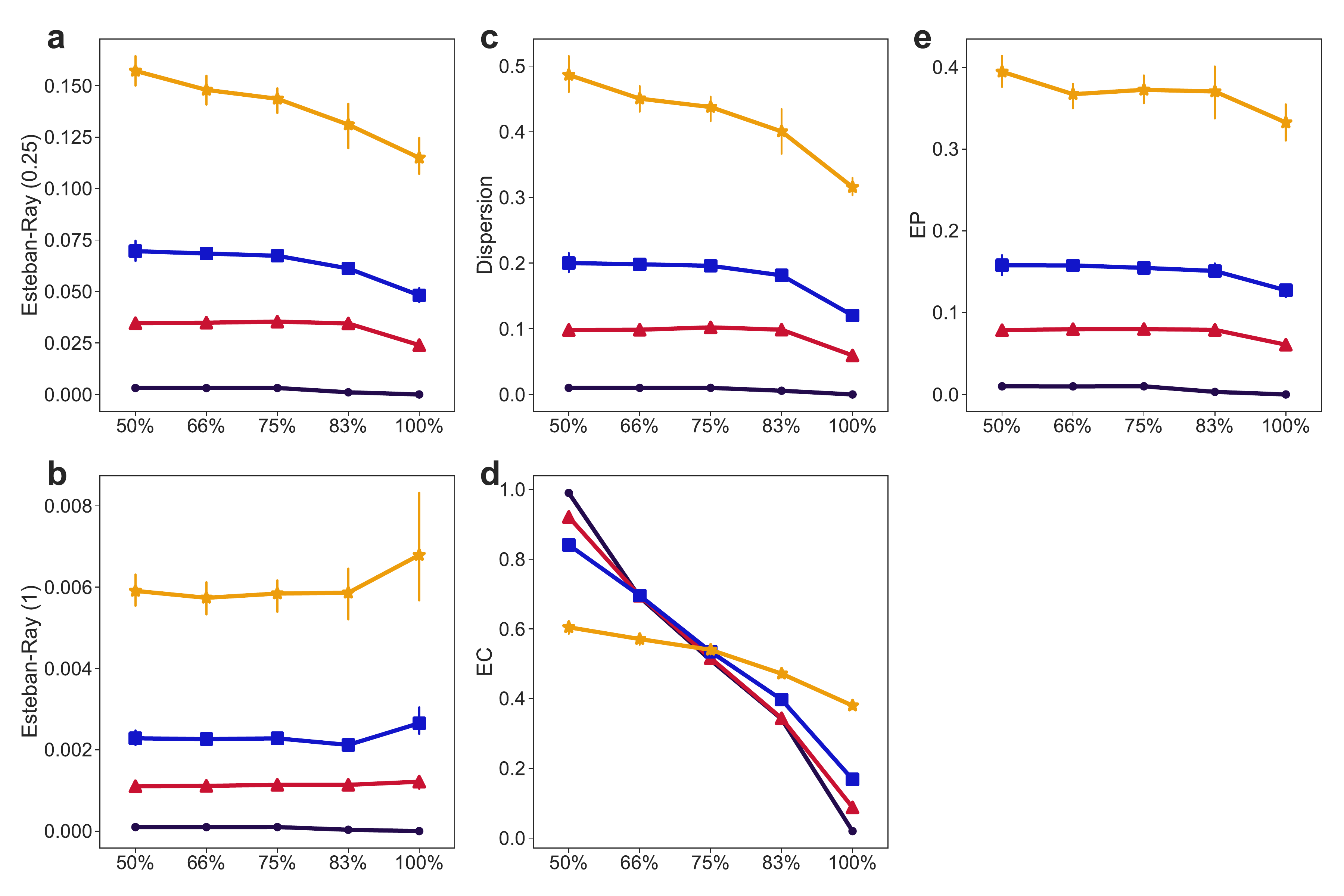}

    \caption{Comparison of traditional polarization measures and EC/EP by using synthetic data. The X-axis represents the sampled mean, and each bar represents a dispersion level. Error bars are calculated by seaborn with 95\% confidence and are, in some places, thinner than the symbols in the figure}
    \label{fig:correlation_polarization}
\end{figure}

\begin{figure}[H]
    \centering
    \includegraphics[width=\columnwidth]{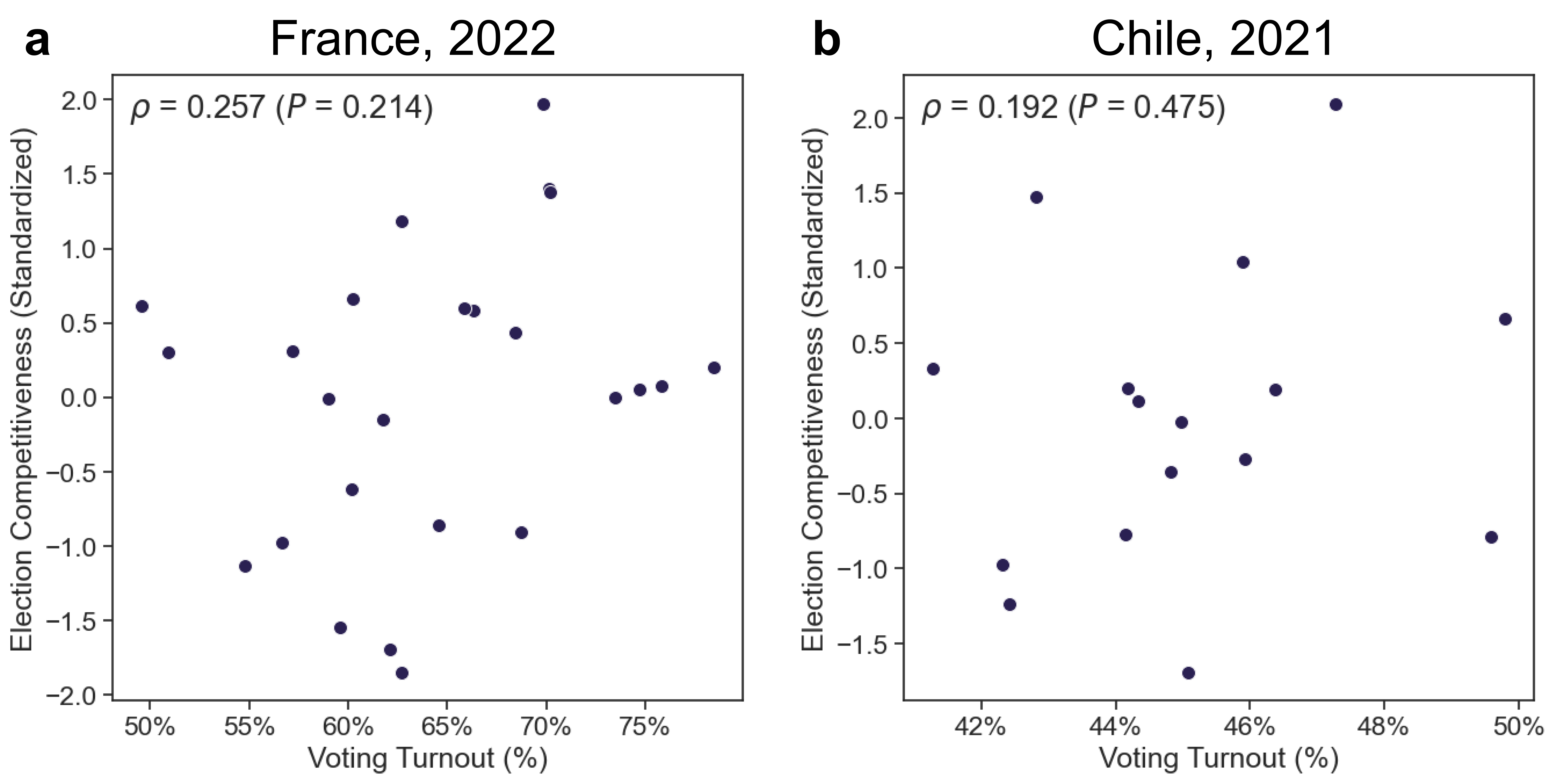}
    \caption{Univariate correlation between EC and Voting Turnout in (a) France, 2022 and (b) Chile, 2021. Both results correspond to the first round.}
    \label{fig:corrturnout}
\end{figure}

We compare those measures both by simulations (Figure \ref{fig:correlation_polarization}) and using data from the last presidential election in Chile (2021) (Table \ref{corrcl}), France (2022) (Table \ref{corrfr}), and the United States (2020) (Table \ref{corrus}).

\subsection{Definition of Swing States}

We define a swing state as one in which there has been at least one winning candidate from both parties (Republican and Democratic) in the last four presidential elections (2008-2020). We label Swing states:  Arizona, Georgia, Indiana, North Carolina, Florida, Iowa, Ohio, Wisconsin, Michigan, and Pennsylvania. Analogously, a Republican state is one in which the Republican Party's candidate won in all four elections; similarly, a Democratic state is one in which the Democratic Party's candidate won in all four elections.

\subsection{Definition of Mass Polarization}

We use the Cooperative Election Study (CES) \autocite{DVN/II2DB6_2022} to estimate Mass Polarization in the United States. The advantage of using CES compared to the American National Election Studies (ANES) is that the first has been previously used to estimate a measure of polarization at the state level. To calculate the ideological score of party $p$ in a state $g$ at a time $t$ as follows:

\begin{align}
    \text{Ideology}_{p,g,t} = \sum\limits_{a=1}^{A}S_{a,p,g,t} \cdot w_{a,g,t}
\end{align}

he parameters include $S_{a,g,t}$ as the ideological score of participant $a$ (1=Lean, 2=Not very strong, 3=Strong) and $w_{a,p,g,t}$ as is the weight of $a$. 

Then, we formally define Mass Polarization (PP) as the perceived ideological distance between the Democratic ($p=1$) and Republican ($p=2$) parties \autocite{baker2020elections}.

\begin{align}
    \text{PP}_{g,t} = |\text{Ideology}_{1,g,t} - \text{Ideology}_{2,g,t}|
\end{align}

From the survey documentation, we use the feature \textbf{pid7}, which collects the 7-point self-reported partisan identity. We numbered that scale in the following form: ``Strong Democrat'': -3, ``Not Very Strong Democrat'': -2, ``Lean Democrat'': -1, ``Independent'': 0, ``Lean Republican'': 1, ``Not Very Strong Republican'': 2, ``Strong Republican'': 3. Values for Not sure were considered as Null.

\subsection{Method Validation}

\subsubsection{Robustness by Voting Percentage}

As mentioned in the main manuscript, we use data from presidential elections in France (2002-2022) and Chile (2013-2021) to test robustness by only using a fraction of the data or only the effective number of candidates given in the election. 

Going back to the manuscript, we first calculate each candidate's voting percentage in the general election. Then we compute the EP by relying on the top-2 candidates, then three candidates, and so on, grouping the excluded candidates under the category ``other''. To perform the analysis, we calculate the similarity (using Pearson correlation) between the EP computed for all candidates and the one calculated for the top-$N$.

\section{Supplementary Results}

\subsection{Regression Analysis}

We further explore the relationship between control variables with respect to EC and EC to analyze whether our measure of polarization satisfies theoretical associations of sociodemographic characteristics with measures of polarization and discards that just mirror inequality or macroeconomic conditions. We set up the regression model as follows:

\begin{align}
y_{g,t} &= \beta_0^ T\mathrm{X}_{g,t} + \mu_t + \gamma_g + \epsilon_{g,t}
\label{eq:antagonism_model}
\end{align}

Where $y_{g,t}$ is the dependent variable for the geography $g$ in year $t$ (in our case, the EP and EC), $\mathrm{X}_{g,t}$ is a vector of control variables in $t$, $\mu_t$ is the period fixed effect, and $\epsilon_{g,t}$ is the error term. We standardized both independent and dependent variables to have a mean of zero and a standard deviation of one. Standard errors are clustered at the state level. Tables \ref{tab:ec} and \ref{tab:ep} use EC and EP as dependent variables, respectively. To complement, Table \ref{tab:md} reports the full coefficient and statistics of models presented in the main manuscript.

\section{Supplementary Tables}

\begin{table}[H]
    \centering
    \begin{tabular}{cp{10cm}c}
    \hline
        Feature & Description & Type \\
        \hline 
        polling\_id & Unique identifier for the minimum aggregation unit & string \\
        candidate & Candidate name & string \\
        value & Candidate's votes in the unit. & integer \\
        rank & Candidate's ranking in the unit. & integer \\
        
        flag\_candidates & Whether to consider candidates are null, abstentions, and blank votes. & boolean \\
        rate & Candidate's percent of votes in the unit & float \\
         \hline
    \end{tabular}
    \caption{Features description of election data sets curated in this manuscript. Each data point represents the results of a candidate in the minimum aggregation unit (e.g., precinct, polling place, county). Each data set is openly shared and deposited in a public repository.}
    \label{tab:features_dataset_1_2}
\end{table}

\begin{table}[H]
    \centering
    \begin{tabular}{cp{10cm}c}
    \hline
        Feature &  Description & type \\
        \hline 
        polling\_id & Unique identifier of the minimum aggregation unit. & string \\
        value & Number of votes of a candidate or party (depending on the election) & integer \\
        rate & Voting percentage of a candidate in the aggregation unit. & float \\
         \hline
    \end{tabular}
    \caption{Features in Data Shared for Location.}
    \label{tab:my_label}
\end{table}

\begin{table}[H]
\begin{center}
\begin{tabular}{l c c c}
\hline
 & Model 1 & Model 2 & Model 3 \\
\hline
Inequality level (t)    & $-0.01$  & $0.03$   & $0.35^{**}$   \\
                        & $(0.22)$ & $(0.24)$ & $(0.17)$      \\
Unemployment (t)        &          & $-0.10$  & $-0.14$       \\
                        &          & $(0.15)$ & $(0.13)$      \\
Ed. Attainment (t)      &          &          & $-0.11$       \\
                        &          &          & $(0.21)$      \\
GDP PC (t)              &          &          & $0.24$        \\
                        &          &          & $(0.23)$      \\
Pop. Density (t)        &          &          & $-0.67^{***}$ \\
                        &          &          & $(0.23)$      \\
\hline
Num. obs.               & $204$    & $204$    & $204$         \\
R$^2$ (full model)      & $0.09$   & $0.10$   & $0.31$        \\
R$^2$ (proj model)      & $0.00$   & $0.00$   & $0.24$        \\
Adj. R$^2$ (full model) & $0.07$   & $0.07$   & $0.28$        \\
Adj. R$^2$ (proj model) & $-0.03$  & $-0.03$  & $0.21$        \\
Num. groups: year       & $4$      & $4$      & $4$           \\
Num. groups: swing      & $2$      & $2$      & $2$           \\
\hline
\multicolumn{4}{l}{\scriptsize{$^{***}p<0.01$; $^{**}p<0.05$; $^{*}p<0.1$}}
\end{tabular}
\caption{State-level association of EP in the United States (2008-2020). Each model represents a multivariate regression, with year-fixed effects. Standard errors are in brackets and are robustly clustered by state. We standardized both independent and dependent variables to have a mean of zero and a standard deviation of one.}
\label{tab:ep}
\end{center}
\end{table}

\begin{table}[H]
\begin{center}
\begin{tabular}{l c c c}
\hline
 & Model 1 & Model 2 & Model 3 \\
\hline
Inequality level (t)    & $-0.38$  & $-0.44$  & $-0.05$       \\
                        & $(0.29)$ & $(0.31)$ & $(0.10)$      \\
Unemployment (t)        &          & $0.18$   & $0.13$        \\
                        &          & $(0.11)$ & $(0.10)$      \\
Ed. Attainment (t)      &          &          & $0.31^{**}$   \\
                        &          &          & $(0.13)$      \\
GDP PC (t)              &          &          & $-0.19$       \\
                        &          &          & $(0.17)$      \\
Pop. Density (t)        &          &          & $-0.68^{***}$ \\
                        &          &          & $(0.12)$      \\
\hline
Num. obs.               & $204$    & $204$    & $204$         \\
R$^2$ (full model)      & $0.23$   & $0.24$   & $0.64$        \\
R$^2$ (proj model)      & $0.16$   & $0.17$   & $0.61$        \\
Adj. R$^2$ (full model) & $0.21$   & $0.22$   & $0.63$        \\
Adj. R$^2$ (proj model) & $0.13$   & $0.15$   & $0.59$        \\
Num. groups: year       & $4$      & $4$      & $4$           \\
Num. groups: swing      & $2$      & $2$      & $2$           \\
\hline
\multicolumn{4}{l}{\scriptsize{$^{***}p<0.01$; $^{**}p<0.05$; $^{*}p<0.1$}}
\end{tabular}
\caption{State-level association of EC in the United States (2008-2020). Each model represents a multivariate regression, with year-fixed effects. Standard errors are in brackets and are robustly clustered by state. We standardized both independent and dependent variables to have a mean of zero and a standard deviation of one.}
\label{tab:ec}
\end{center}
\end{table}

\begin{table}[H]
\begin{center}
\begin{tabular}{l c c c c}
\hline
 & MP (1) & MP (2) & VT (1) & VT (2) \\
\hline
EP (t)                  & $0.33^{***}$ & $0.31^{***}$  & $-0.01$     & $0.29^{**}$  \\
                        & $(0.07)$     & $(0.09)$      & $(0.13)$    & $(0.14)$     \\
EC (t)                  & $-0.11^{*}$  & $0.04$        & $0.21^{**}$ & $0.55^{***}$ \\
                        & $(0.06)$     & $(0.10)$      & $(0.10)$    & $(0.18)$     \\
Inequality level (t)    &              & $0.26^{**}$   &             & $-0.27^{**}$ \\
                        &              & $(0.10)$      &             & $(0.11)$     \\
Unemployment (t)        &              & $0.01$        &             & $-0.05$      \\
                        &              & $(0.11)$      &             & $(0.06)$     \\
Ed. Attainment (t)      &              & $-0.36^{***}$ &             & $0.53^{***}$ \\
                        &              & $(0.13)$      &             & $(0.11)$     \\
GDP PC (t)              &              & $-0.10$       &             & $0.03$       \\
                        &              & $(0.19)$      &             & $(0.16)$     \\
Pop. Density (t)        &              & $0.28$        &             & $0.31$       \\
                        &              & $(0.17)$      &             & $(0.23)$     \\
\hline
Num. obs.               & $204$        & $204$         & $303$       & $201$        \\
R$^2$ (full model)      & $0.47$       & $0.60$        & $0.30$      & $0.62$       \\
R$^2$ (proj model)      & $0.17$       & $0.38$        & $0.05$      & $0.51$       \\
Adj. R$^2$ (full model) & $0.45$       & $0.58$        & $0.28$      & $0.60$       \\
Adj. R$^2$ (proj model) & $0.15$       & $0.34$        & $0.03$      & $0.48$       \\
Num. groups: year       & $4$          & $4$           & $6$         & $4$          \\
Num. groups: swing      & $2$          & $2$           & $2$         & $2$          \\
\hline
\multicolumn{5}{l}{\scriptsize{$^{***}p<0.01$; $^{**}p<0.05$; $^{*}p<0.1$}}
\end{tabular}
\caption{State-level association for Mass Polarization (MP) and Voting Turnout (VT) in the United States (2008-2020). Each model represents a multivariate regression, with year-fixed effects. These models represent the ones presented in the main manuscript. Standard errors are in brackets and are robustly clustered by state. We standardized both independent and dependent variables to have a mean of zero and a standard deviation of one.}
\label{tab:md}
\end{center}
\end{table}

\begin{table}[H]
\begin{center}
\begin{tabular}{lccccc}
\hline
Method & (1) & (2) & (3) & (4) & (5) \\
\hline
(1) EP & 1.0 & -0.24 & 0.03 & -0.29 & 0.99 \\
(2) EC & - & 1.0 & -0.21 & -0.1 & -0.25 \\
(3) Esteban-Ray (0.25) & - & - & 1.0 & 0.85 & 0.12 \\
(4) Esteban-Ray (1) & - & - & - & 1.0 & -0.21 \\
(5) Dispersion & - & - & - & - & 1.0 \\
\hline
\end{tabular}
\caption{Correlation matrix for measures of polarization in Chile (2021)}
\label{corrcl}
\end{center}
\end{table}

\begin{table}[H]
\begin{center}
\begin{tabular}{lccccc}
\hline
Method & (1) & (2) & (3) & (4) & (5) \\
\hline
(1) EP & 1.0 & 0.38 & -0.04 & -0.33 & 0.67 \\
(2) EC & - & 1.0 & -0.51 & -0.71 & 0.33 \\
(3) Esteban-Ray (0.25) & - & - & 1.0 & 0.86 & 0.32 \\
(4) Esteban-Ray (1) & - & - & - & 1.0 & -0.06 \\
(5) Dispersion & - & - & - & - & 1.0 \\
\hline
\end{tabular}
\caption{Correlation matrix for measures of polarization in France (2022)}
\label{corrfr}
\end{center}
\end{table}

\begin{table}[H]
\begin{center}
\begin{tabular}{lccccc}
\hline
Method & (1) & (2) & (3) & (4) & (5) \\
\hline
(1) EP & 1.0 & -0.55 & 0.82 & 0.03 & 0.5 \\
(2) EC & - & 1.0 & -0.46 & 0.05 & -0.18 \\
(3) Esteban-Ray (0.25) & - & - & 1.0 & 0.47 & 0.49 \\
(4) Esteban-Ray (1) & - & - & - & 1.0 & 0.06 \\
(5) Dispersion & - & - & - & - & 1.0 \\
\hline
\end{tabular}
\caption{Correlation matrix for measures of polarization in the United States (2020)}
\label{corrus}
\end{center}
\end{table}

\section{Supplementary Figures}

\begin{figure}[H]
    \centering
    \includegraphics[width=\columnwidth]{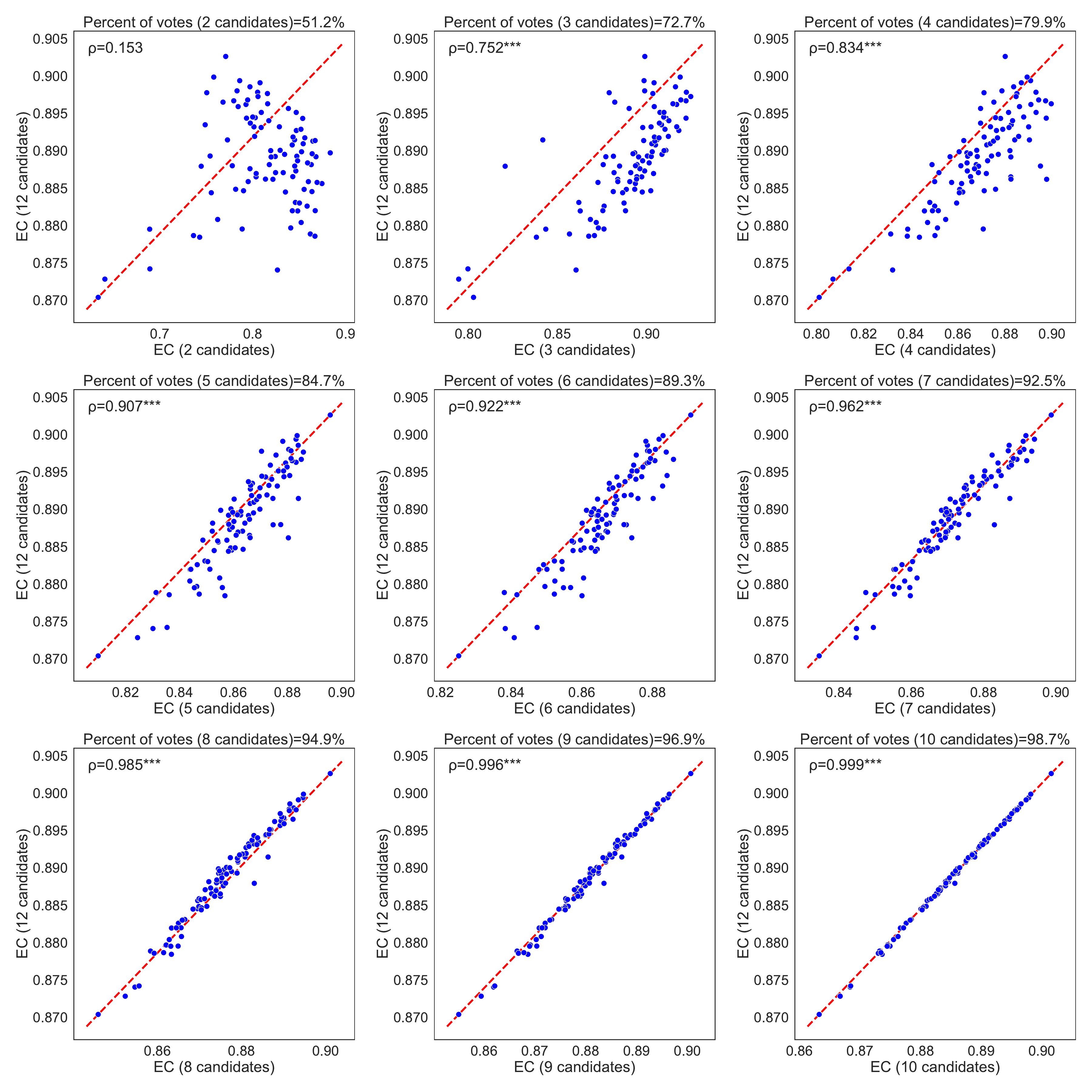}

    \caption{Region-level EC robustness in France (2022) by excluding candidates from the analysis. Note: ***$p<0.01$, **$p<0.05$, *$p<0.1$.}
    \label{fig:rb_a_fr_2022}
\end{figure}

\begin{figure}[H]
    \centering
    \includegraphics[width=\columnwidth]{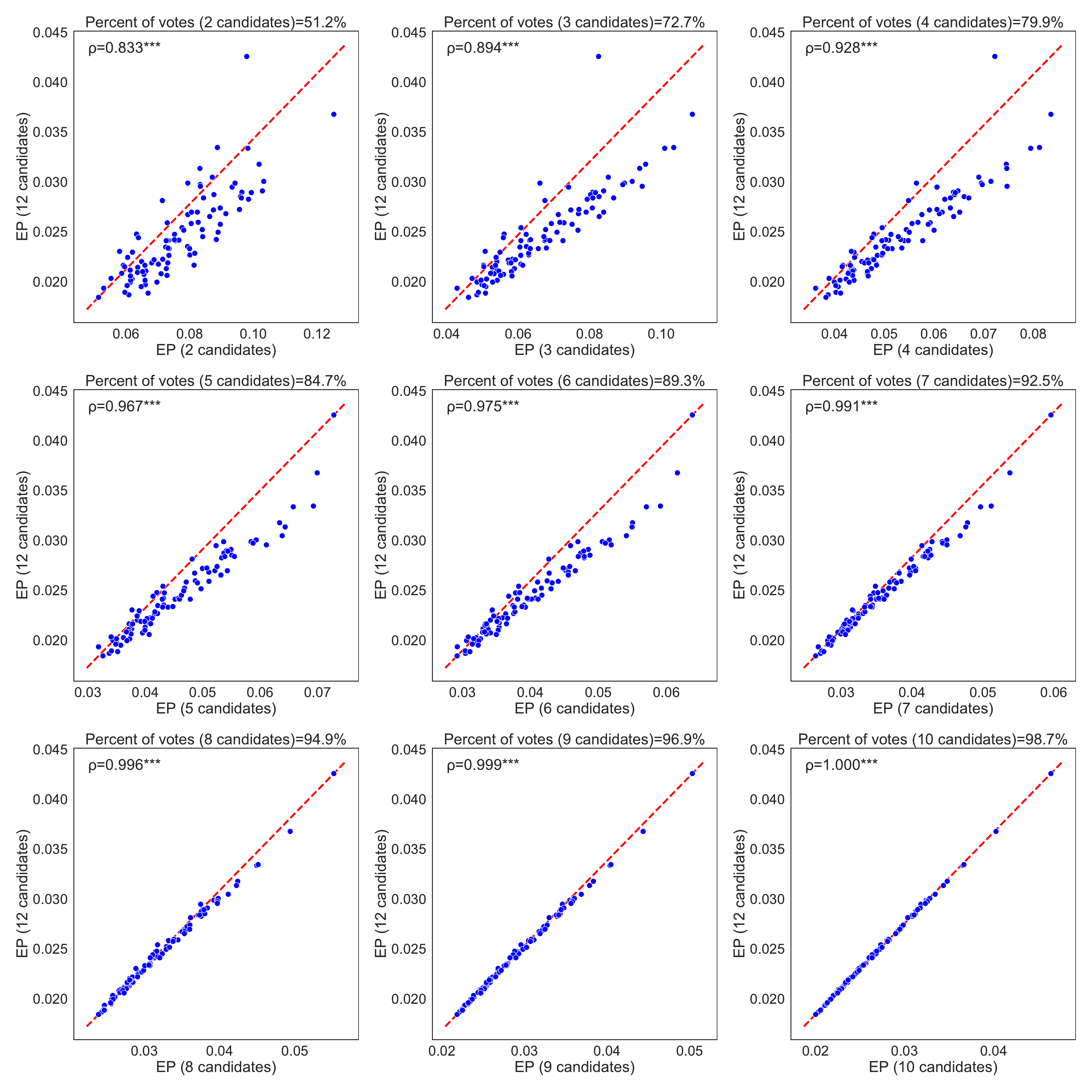}

    \caption{Region-level EP robustness in France (2022) by excluding candidates from the analysis. Note: ***$p<0.01$, **$p<0.05$, *$p<0.1$.}
    \label{fig:rb_b_fr_2022}
\end{figure}

\begin{figure}[H]
    \centering
    \includegraphics[width=\columnwidth]{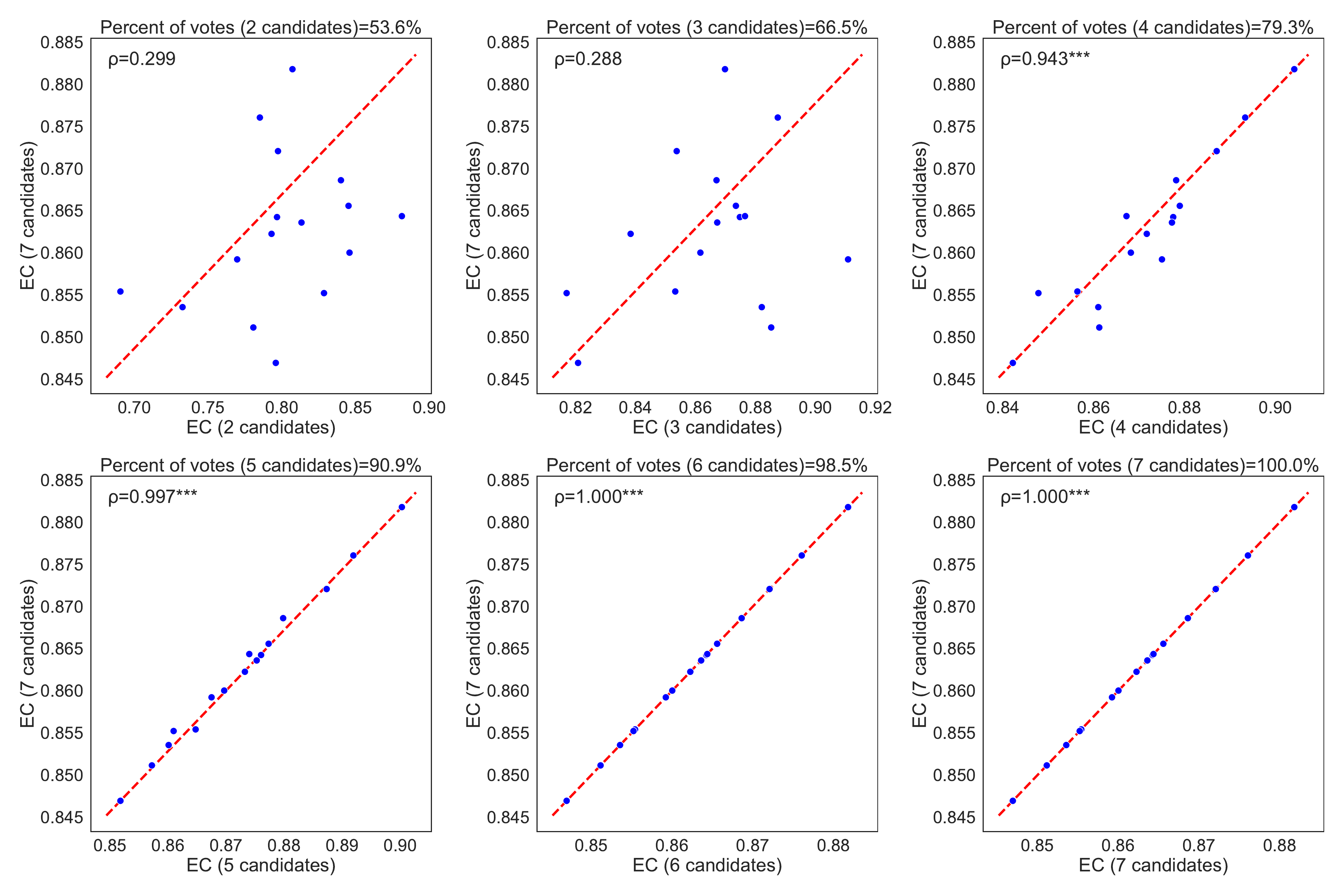}

    \caption{Region-level EC robustness in Chile (2021) by excluding candidates from the analysis. Note: ***$p<0.01$, **$p<0.05$, *$p<0.1$.}
    \label{fig:rb_a_cl_2021}
\end{figure}

\begin{figure}[H]
    \centering
    \includegraphics[width=\columnwidth]{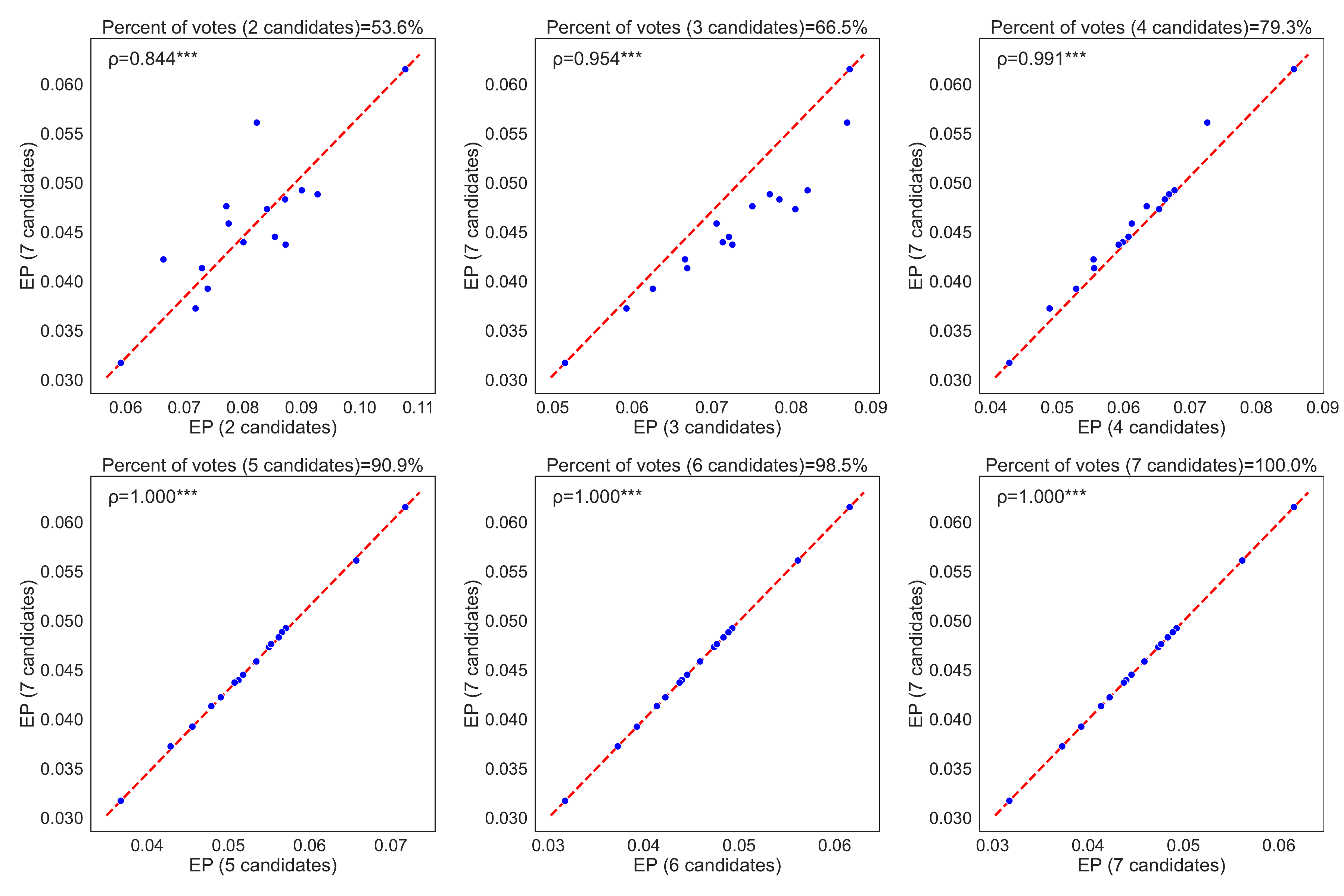}

    \caption{Region-level EP robustness in Chile (2021) by excluding candidates from the analysis. Note: ***$p<0.01$, **$p<0.05$, *$p<0.1$.}
    \label{fig:rb_b_cl_2021}
\end{figure}

% \begin{figure}[H]
%     \centering
%     \includegraphics[width=\columnwidth]{Figures_SM/Comparison_First_Round_Runoff.pdf}
%     \caption{Comparison of EP and EC between First Round and Runoff \textit{Note: *p$<$0.1; **p$<$0.05; ***p$<$0.01}}
%     \label{fig:comparison_first_runoff}
% \end{figure}

% \bibliographystyle{naturemag} % a naturemag We choose the "plain" reference style
% \bibliography{SupRefs}% common bib file
\printbibliography[heading=bibintoc]

\end{refsection}
\end{document}